\documentclass{sig-alternate-2013}

\newfont{\mycrnotice}{ptmr8t at 7pt}
\newfont{\myconfname}{ptmri8t at 7pt}
\let\crnotice\mycrnotice%
\let\confname\myconfname%

\usepackage{algorithm}
\usepackage{algorithmic}
\usepackage[caption=false,font=footnotesize]{subfig}
\usepackage{amsmath}
\usepackage{comment}
\usepackage{listings}
\usepackage{multirow}
\usepackage[usenames,dvipsnames]{color}
\usepackage{footnote}

\newcommand{\Paragraph}[1]{\smallskip\noindent{\bf #1.}}

\begin{document}



\title{\Large \bf i$^2$MapReduce: Incremental MapReduce for Mining Evolving Big Data}

\numberofauthors{4} 
%
\author{
%
%
\alignauthor
Yanfeng Zhang\\
       \affaddr{Northeastern University, China}\\
       \email{zhangyf@cc.neu.edu.cn}
\alignauthor
Shimin Chen\\
       \affaddr{Institute of Computing Technology, CAS}\\
       \email{chensm@ict.ac.cn}
\alignauthor
Qiang Wang\\
       \affaddr{Northeastern University, China}\\
       \email{403268229@qq.com}
\and  
\alignauthor
Ge Yu\\
       \affaddr{Northeastern University, China}\\
       \email{yuge@mail.neu.edu.cn}
}

\maketitle

\begin{abstract}
As new data and updates are constantly arriving, the results of data
mining applications become stale and obsolete over time.  Incremental
processing is a promising approach to refreshing mining results.
It utilizes previously saved states to avoid the expense of
re-computation from scratch.

In this paper, we propose i$^2$MapReduce, a novel incremental
processing extension to MapReduce, the most widely used framework for
mining big data.  Compared with the state-of-the-art work on Incoop,
i$^2$MapReduce (i) performs key-value pair level incremental
processing rather than task level re-computation, (ii) supports not
only one-step computation but also more sophisticated iterative
computation, which is widely used in data mining applications, and
(iii) incorporates a set of novel techniques to reduce I/O overhead
for accessing preserved fine-grain computation states.  We evaluate
i$^2$MapReduce using a one-step algorithm and three iterative
algorithms with diverse computation characteristics.  Experimental
results on Amazon EC2 show significant performance improvements of
i$^2$MapReduce compared to both plain and iterative MapReduce
performing re-computation.
\end{abstract}

\section{Introduction}
\label{sec:intro}

Today huge amount of digital data is being accumulated in many
important areas, including e-commerce, social network, finance, health
care, education, and environment.  It has become increasingly popular
to mine such big data in order to gain insights to help business
decisions or to provide better personalized, higher quality services.
In recent years, a large number of computing
frameworks~\cite{1251264,sparknsdi,piccolo,1807184,rex,distgraphlab,Ewen:2012:SFI:2350229.2350245,haloop,
Ekanayake:2010:TRI:1851476.1851593, imapreduce} have been developed
for big data analysis.  Among these frameworks,
MapReduce~\cite{1251264} (with its open-source implementations, such
as Hadoop) is the most widely used in production because of its
simplicity, generality, and maturity.  We focus on improving MapReduce
in this paper.

Big data is constantly evolving.  As new data and updates are being
collected, the input data of a big data mining algorithm will
gradually change, and the computed results will become stale and
obsolete over time.  In many situations, it is desirable to
periodically refresh the mining computation in order to keep the
mining results up-to-date.  For example, the PageRank
algorithm~\cite{Brin:1998:ALH:297810.297827} computes ranking scores
of web pages based on the web graph structure for supporting web
search.  However, the web graph structure is constantly evolving; Web
pages and hyper-links are created, deleted, and updated.  As the
underlying web graph evolves, the PageRank ranking results gradually
become stale, potentially lowering the quality of web search.
Therefore, it is desirable to refresh the PageRank computation
regularly.

Incremental processing is a promising approach to refreshing mining
results.  Given the size of the input big data, it is often very
expensive to rerun the entire computation from scratch.
Incremental processing exploits the fact that the input data of two
subsequent computations A and B are similar.  Only a very small
fraction of the input data has changed.  The idea is to save states in
computation A, re-use A's states in computation B, and perform
re-computation only for states that are affected by the changed input
data.  In this paper, we investigate the realization of this principle
in the context of the MapReduce computing framework.

A number of previous studies (including
Percolator~\cite{Peng:2010:LIP:1924943.1924961},
CBP~\cite{Logothetis:2010:SBP:1807128.1807138}, and
Naiad~\cite{Murray:2013:NTD:2517349.2522738}) have followed this
principle and designed new programming models to support incremental
processing.  Unfortunately, the new programming models (BigTable
observers in Percolator, stateful translate operators in CBP, and
timely dataflow paradigm in Naiad) are drastically different from
MapReduce, requiring programmers to completely re-implement
their algorithms.

On the other hand, Incoop~\cite{incoop} extends MapReduce to support
incremental processing.  However, it has two main limitations.
First, Incoop supports only \emph{task-level} incremental processing.
That is, it saves and reuses states at the granularity of individual
Map and Reduce tasks.  Each task typically processes a large number of
key-value pairs (kv-pairs). If Incoop detects any data changes in the
input of a task, it will rerun the entire task.  While this approach
easily leverages existing MapReduce features for state savings, it may
incur a large amount of redundant computation if only a small fraction
of kv-pairs have changed in a task.
Second, Incoop supports only \emph{one-step} computation,  while
important mining algorithms, such as PageRank, require iterative
computation.  Incoop would treat each iteration as a separate
MapReduce job.  However, a small number of input data changes may
gradually propagate to affect a large portion of intermediate states
after a number of iterations, resulting in expensive global
re-computation afterwards.

We propose i$^2$MapReduce, an extension to MapReduce that supports
\emph{fine-grain} incremental processing for both \emph{one-step} and
\emph{iterative} computation.  Compared to previous solutions,
i$^2$MapReduce incorporates the following three novel features:

\begin{list}{\labelitemi}{\setlength{\leftmargin}{3mm}\setlength{\itemindent}{-1mm}\setlength{\topsep}{0mm}\setlength{\itemsep}{0mm}\setlength{\parsep}{1mm}}

\item \textbf{Fine-grain Incremental Processing using MRBG-Store}:
Unlike Incoop, i$^2$MapReduce supports kv-pair level fine-grain
incremental processing in order to minimize the amount of
re-computation as much as possible.  We model the kv-pair level data
flow and data dependence in a MapReduce computation as a bipartite
graph, called MRBGraph.  A MRBG-Store is designed to preserve the
fine-grain states in the MRBGraph and support efficient queries to
retrieve fine-grain states for incremental processing. (cf.
Section~\ref{sec:incr})

\item \textbf{General-Purpose Iterative Computation with Modest
Extension to MapReduce API}: Our previous work proposed
iMapReduce~\cite{imapreduce} to efficiently support iterative
computation on the MapReduce platform. However, it targets types of iterative computation where there is a one-to-one/all-to-one correspondence from Reduce output to Map input.
In comparison, our
current proposal provides general-purpose support, including not only
one-to-one, but also one-to-many, many-to-one, and many-to-many
correspondence.  We enhance the Map API to allow users to easily
express loop-invariant structure data, and we propose a Project API
function to express the correspondence from Reduce to Map.  While
users need to slightly modify their algorithms in order to take full
advantage of i$^2$MapReduce, such modification is modest compared to
the effort to re-implement algorithms on a completely different
programming paradigm, such as in
Percolator~\cite{Peng:2010:LIP:1924943.1924961},
CBP~\cite{Logothetis:2010:SBP:1807128.1807138}, and
Naiad~\cite{Murray:2013:NTD:2517349.2522738}. (cf.
Section~\ref{sec:iter})

\item \textbf{Incremental Processing for Iterative Computation}:
Incremental iterative processing is substantially more challenging
than incremental one-step processing because even a small number of
updates may propagate to affect a large portion of intermediate states
after a number of iterations.  To address this problem, we propose to
reuse the converged state from the previous computation and employ a
change propagation control mechanism.  We also enhance the MRBG-Store
to better support the access patterns in incremental iterative
processing.  To our knowledge, i$^2$MapReduce is the \emph{first
MapReduce-based solution that efficiently supports incremental
iterative computation}. (cf. Section~\ref{sec:incriter})

\end{list}

\noindent We implemented i$^2$MapReduce by modifying Hadoop-1.0.3.
We evaluate i$^2$MapReduce using a one-step algorithm
(A-Priori) and three iterative algorithms (PageRank, Kmeans, GIM-V)
with diverse computation characteristics. Experimental results on
Amazon EC2 show significant performance improvements of i$^2$MapReduce
compared to both plain and iterative MapReduce performing
re-computation.  For example, for the iterative PageRank computation
with 10\% data changed, i$^2$MapReduce improves the run time of
re-computation on plain MapReduce by a 8 fold speedup. (cf.
Section~\ref{sec:expr})

\section{MapReduce Background}
\label{sec:background}

\begin{figure}[t]
   \centering
      \includegraphics[width=2.9in]{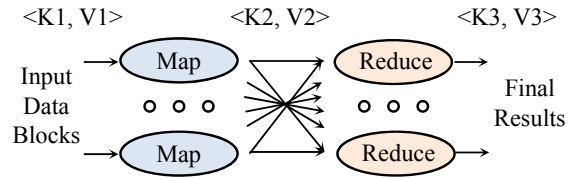}
   \caption{MapReduce computation.}
   \label{fig:mr}
\end{figure}

A MapReduce program is composed of a Map function and a
Reduce function~\cite{1251264}, as shown in Fig.~\ref{fig:mr}. Their
APIs are as follows:

\begin{equation}\label{eq:orig-map}
  \texttt{map}(K1,V1)\rightarrow [\langle K2,V2\rangle]\notag
\end{equation}

\begin{equation}\label{eq:orig-reduce}
  \texttt{reduce}(K2,\{V2\})\rightarrow [\langle K3,V3\rangle]\notag
\end{equation}

\noindent The Map function takes a kv-pair $\langle K1,
V1\rangle$ as input and computes zero or more intermediate kv-pairs
$\langle K2, V2\rangle$s.  Then all $\langle K2, V2\rangle$s are
grouped by $K2$.  The Reduce function takes a $K2$ and a list of
$\{V2\}$ as input and computes the final output kv-pairs $\langle K3,
V3\rangle$s.

A MapReduce system (e.g., Apache Hadoop) usually reads the input data
of the MapReduce computation from and writes the final results to a
distributed file system (e.g., HDFS), which divides a file into
equal-sized (e.g., 64MB) blocks and stores the blocks across a cluster
of machines.  For a MapReduce program, the MapReduce system runs a
JobTracker process on a master node to monitor the job progress, and a
set of TaskTracker processes on worker nodes to perform the actual Map
and Reduce tasks.

The JobTracker starts a Map task per data block, and typically assigns
it to the TaskTracker on the machine that holds the corresponding data
block in order to minimize communication overhead.  Each Map task
calls the Map function for every input $\langle K1, V1\rangle$, and
stores the intermediate kv-pairs $\langle K2, V2\rangle$s on local
disks.  Intermediate results are shuffled to Reduce tasks according to
a partition function (e.g., a hash function) on $K2$.  After a Reduce
task obtains and merges intermediate results from all Map Tasks, it
invokes the Reduce function on each $\langle K2, \{V2\}\rangle$ to
generate the final output kv-pairs $\langle K3, V3\rangle$s.

\section{Fine-grain Incremental Processing for One-Step Computation}
\label{sec:incr}

We begin by describing the basic idea of fine-grain incremental
processing in Section~\ref{subsec:idea}.  In
Section~\ref{subsec:mrgraph}--\ref{subsec:engine}, we present the main
design, including the MRBGraph abstraction and the incremental
processing engine.  Then in
Section~\ref{subsec:mrbgstore}--\ref{subsec:accumulator}, we delve
into two aspects of the design, i.e. the mechanism that preserves the
fine-grain states, and the handling of a special but popular case
where the Reduce function performs accumulation operations.

\subsection{Basic Idea}
\label{subsec:idea}

Consider two MapReduce jobs $A$ and $A'$ performing the same
computation on input data set $D$ and $D'$, respectively.  $D'= D +
\Delta D$, where $\Delta D$ consists of the inserted and deleted input
$\langle K1, V1\rangle$s\footnote{We assume that new data or new
updates are captured via incremental data acquisition or incremental
crawling~\cite{ilprints376,olston2010web}.  Incremental data
acquisition can significantly save the resources for data collection;
it does not re-capture the whole data set but only capture the
revisions since the last time that data was captured.}.  An update can
be represented as a deletion followed by an insertion.  Our goal is to
re-compute only the Map and Reduce function call instances that are
affected by $\Delta D$.

Incremental computation for Map is straightforward. We simply invoke
the Map function for the inserted or deleted $\langle K1, V1\rangle$s.
Since the other input kv-pairs are not changed, their Map computation
would remain the same.  We now have computed the delta intermediate
values, denoted $\Delta M$, including inserted and deleted $\langle
K2, V2\rangle$s.

To perform incremental Reduce computation, we need to save the
fine-grain states of job $A$, denoted $M$, which includes $\langle K2, \{V2\}\rangle$s.
We will re-compute the Reduce function for each $K2$ in $\Delta M$.
The other $K2$ in $M$ does not see any changed intermediate values and
therefore would generate the same final result.  For a $K2$ in $\Delta
M$, typically only a subset of the list of $V2$ have changed.  Here,
we retrieve the saved $\langle K2, \{V2\}\rangle$ from $M$, and
apply the inserted and/or deleted values from $\Delta M$ to obtain an
updated Reduce input.  We then re-compute the Reduce function on this
input to generate the changed final results $\langle K3, V3\rangle$s.

It is easy to see that results generated from this incremental
computation are logically the same as the results from completely
re-computing $A'$.

\subsection{MRBGraph Abstraction}
\label{subsec:mrgraph}

\begin{figure}[t]
    \centering
        \includegraphics[width=2.9in]{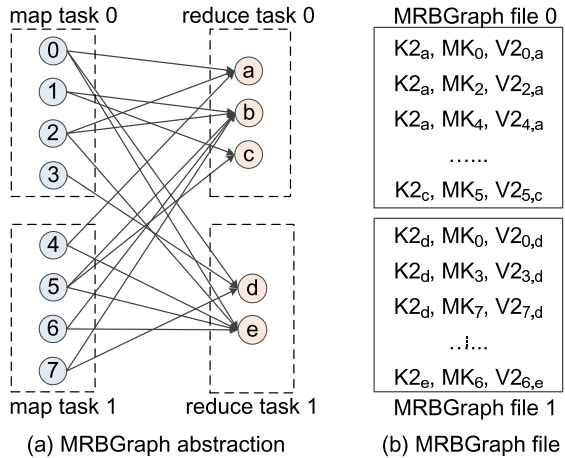}
    \caption{MRBGraph.}
    \label{fig:mrbgraph}
\end{figure}

We use a MRBGraph (Map Reduce Bipartite Graph) abstraction to model
the data flow in MapReduce, as shown in Fig.~\ref{fig:mrbgraph} (a).
Each vertex in the Map task represents an individual
Map function call instance on a pair of $\langle K1, V1\rangle$. Each
vertex in the Reduce task represents an individual Reduce function
call instance on a group of $\langle K2,\{V2\}\rangle$.  An edge from
a Map instance to a Reduce instance means that the Map instance
generates a $\langle K2,V2\rangle$ that is shuffled to become part of
the input to the Reduce instance.  For example, the input of Reduce
instance $a$ comes from Map instance $0$, $2$, and $4$.

MRBGraph edges are the fine-grain states $M$ that we would like to
preserve for incremental processing.  An edge contains three pieces of
information: (i) the source Map instance, (ii) the destination Reduce
instance (as identified by $K2$), and (iii) the edge value (i.e.
$V2$).  Since Map input key $K1$ may not be unique, i$^2$MapReduce
generates a globally unique Map key $MK$ for each Map instance.
Therefore, i$^2$MapReduce will preserve ($K2$, $MK$, $V2$) for each
MRBGraph edge.

\subsection{Fine-grain Incremental Processing Engine}
\label{subsec:engine}

\begin{figure}[t]
    \centering
        \includegraphics[width=3.1in]{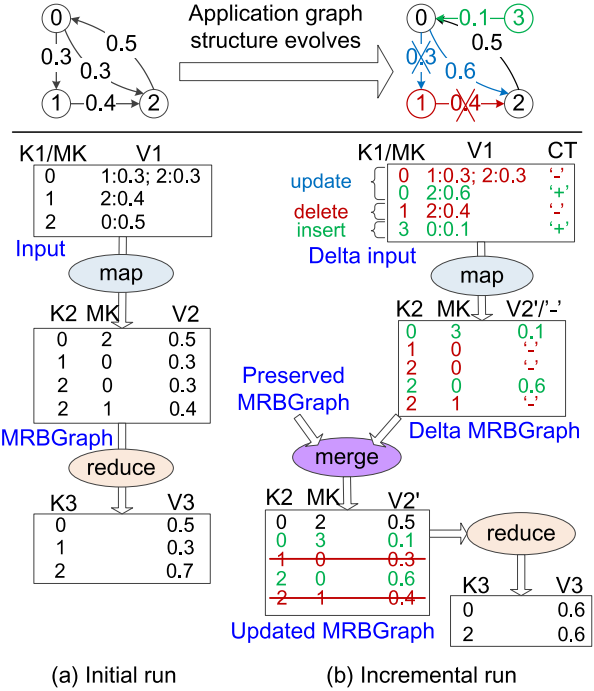}
    \caption{Incremental processing for an application that computes the sum of in-edge weights for each vertex.}
    \label{fig:increxam}
\end{figure}

Fig.~\ref{fig:increxam} illustrates the fine-grain incremental
processing engine with an example application, which computes the sum
of in-edge weights for each vertex in a graph.  As shown at the top of
Fig.~\ref{fig:increxam}, the input data, i.e. the graph structure,
evolves over time.  In the following, we describe how the engine
performs incremental processing to refresh the analysis results.

\Paragraph{Initial Run and MRBGraph Preserving}  The initial run
performs a normal MapReduce job, as shown in Fig.~\ref{fig:increxam}
(a).  The Map input is the adjacency matrix of the graph.  Every
record corresponds to a vertex in the graph.  $K1$ is vertex id $i$,
and $V1$ contains ``$j_1$:$w_{i,j_1}$; $j_2$:$w_{i,j_2}$; \ldots''
where $j$ is a destination vertex and $w_{i,j}$ is the weight of the
out-edge $(i,j)$.  Given such a record, the Map function outputs
intermediate kv-pair $\langle j, w_{i,j}\rangle$ for every $j$.  The
shuffling phase groups the edge weights by the destination vertex.
Then the Reduce function computes for a vertex $j$ the sum of all its
in-edge weights as $\sum_{i}w_{i,j}$.

For incremental processing, we preserve the fine-grain MRBGraph edge
states.  A question arises: shall the states be preserved at the Map
side or at the Reduce side? We choose the latter because during
incremental processing original intermediate values can be obtained at
the Reduce side without any shuffling overhead.  The engine transfers
the globally unique $MK$ along with $\langle K2, V2\rangle$ during the
shuffle phase.  Then it saves the states ($K2, MK, V2$) in a MRBGraph
file at every Reduce task, as shown in Fig.~\ref{fig:mrbgraph} (b).

\Paragraph{Delta Input}  i$^2$MapReduce expects delta input data that
contains the newly inserted, deleted, or modified kv-pairs as the
input to incremental processing.  Note that identifying the data
changes is beyond the scope of this paper; Many incremental data
acquisition or incremental crawling techniques have been developed to
improve data collection performance~\cite{ilprints376,olston2010web}.

Fig.~\ref{fig:increxam} (b) shows the delta input for the updated
application graph.  A `+' symbol indicates a newly inserted kv-pair,
while a `-' symbol indicates a deleted kv-pair.  An update is represented as a
deletion followed by an insertion.  For example, the deletion of vertex
$1$ and its edge are reflected as $\langle 1, 2$:$0.4,$`-'$\rangle$.  The
insertion of vertex $3$ and its edge leads to $\langle 3,
0$:$0.1,$`+'$\rangle$.
The modification of the vertex $0$'s edges are reflected by a
deletion of the old record $\langle 0, 1$:$0.3$;$2$:$0.3,$`-'$\rangle$
and an insertion of a new record $\langle 0, 2$:$0.6,$`-'$\rangle$.

\Paragraph{Incremental Map Computation to Obtain the Delta MRBGraph}
The engine invokes the Map function for every record in the delta
input.  For an insertion with `+', its intermediate results $\langle
K2,MK,V2'\rangle$s represent newly inserted edges in the MRBGraph.
For a deletion with `-', its intermediate results indicate that the
corresponding edges have been removed from the MRBGraph.  The engine
replaces the $V2'$s of the deleted MRBGraph edges with `-'.  During
the MapReduce shuffle phase, the intermediate $\langle
K2,MK,V2'\rangle$s and $\langle K2,MK,$`-'$\rangle$s with the same $K2$
will be grouped together.  The delta MRBGraph will contain only the
changes to the MRBGraph and sorted by the $K2$ order.

\Paragraph{Incremental Reduce Computation}  The engine merges the
delta MRBGraph and the preserved MRBGraph to obtain the updated
MRBGraph using the algorithm in
Section~\ref{subsec:mrbgstore}.  For each $\langle K2,MK,$`-'$\rangle$,
the engine deletes the corresponding saved edge state.  For each
$\langle K2,MK,V2'\rangle$, the engine first checks duplicates, and inserts the new edge if no duplicate exists, or else updates the old edge if duplicate exists.
(Note that ($K2$,$MK$) uniquely identifies a MRBGraph edge.)
Since an update in the Map input is represented as a deletion and an
insertion, any modification to the intermediate edge state (e.g.,
$\langle 2,0,* \rangle$ in the example) consists of a deletion (e.g.,
$\langle 2,0,$`-'$\rangle$) followed by an insertion (e.g., $\langle
2,0,0.6 \rangle$).  For each affected $K2$,  the merged list of $V2$
will be used as input to invoke the Reduce function to generate the
updated final results.

\subsection{MRBG-Store}
\label{subsec:mrbgstore}

The MRBG-Store supports the preservation and retrieval of fine-grain
MRBGraph states for incremental processing.  We see two main
requirements on the MRBG-Store.
First, the MRBG-Store must incrementally store the evolving MRBGraph.
Consider a sequence of jobs that incrementally refresh the results of
a big data mining algorithm.  As input data evolves, the intermediate
states in the MRBGraph will also evolve.  It would be wasteful to
store the entire MRBGraph of each subsequent job.  Instead, we would
like to obtain and store only the updated part of the MRBGraph.
Second, the MRGB-Store must support efficient retrieval of preserved
states of given Reduce instances.   For incremental Reduce
computation, i$^2$MapReduce re-computes the Reduce instance associated
with each changed MRBGraph edge, as described in
Section~\ref{subsec:engine}.  For a changed edge, it queries the
MRGB-Store to retrieve the preserved states of the in-edges of the
associated $K2$, and merge the preserved states with the newly
computed edge changes.

\begin{figure}[t]
    \centering
        \includegraphics[width=3.2in]{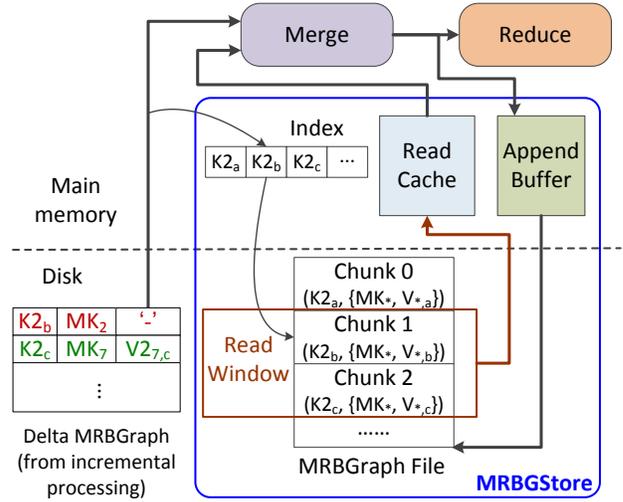}
    \caption{Structure of MRBG-Store.}
    \label{fig:mrbgstore}
\end{figure}

Fig.~\ref{fig:mrbgstore} depicts the structure of the MRBG-Store.
We describe how the components of the MRBG-Store work together to
achieve the above two requirements.

\Paragraph{Fine-grain State Retrieval and Merging} A MRBGraph file
stores fine-grain intermediate states for a Reduce task, as
illustrated previously in Fig.~\ref{fig:mrbgraph} (b).  In
Fig.~\ref{fig:mrbgstore}, we see that the $\langle K2,MK,V2\rangle$s
with the same $K2$ are stored contiguously as a \emph{chunk}.  Since a
chunk corresponds to the input to a Reduce instance, our design treats
chunk as the basic unit, and always reads, writes, and operates on
entire chunks.

The contents of a delta MRBGraph file are shown on the bottom left of
Fig.~\ref{fig:mrbgstore}.  Every record represents a change in the
original (last preserved) MRBGraph.  There are two kinds of records.
An edge insertion record (in green color) contains a valid $V2$ value;
an edge deletion record (in red color) contains a null value (as
marked by `-').

\renewcommand{\algorithmicrequire}{\textbf{Input}}
\renewcommand{\algorithmicensure}{\textbf{Output}}

\begin{algorithm}[t]
\caption{Query Algorithm in MRBG-Store}
\label{alg:query}
\begin{algorithmic}[1]
\REQUIRE queried key: $k$; the list of queried keys: $L$
\ENSURE chunk $k$
\IF{! $read\_cache$.contains($k$)}
\STATE $gap\leftarrow 0$, $w\leftarrow 0$
\STATE $i\leftarrow$ $k$'s index in $L$   \quad\quad\quad\quad  $//$ That is, $L_i=k$
\WHILE{$gap<T$ and $w+gap+length(L_{i})<read\_cache.size$}
\STATE $w\leftarrow w+gap+length(L_{i})$
\STATE $gap\leftarrow pos(L_{i+1})-pos(L_i)-length(L_i)$
\STATE $i\leftarrow i+1$
\ENDWHILE
\STATE starting from $pos(k)$, read $w$ bytes into $read\_cache$
\ENDIF
\STATE return $read\_cache$.get\_chunk($k$)
\end{algorithmic}
\end{algorithm}

The merging of the delta MRBGraph with the MRBGraph file in the
MRBG-Store is essentially a join operation using $K2$ as the join key.
Since the size of the delta MRBGraph is typically much smaller than
the MRBGraph file, it is wasteful to read the entire MRBGraph file.
Therefore, we construct an index for selective access to the MRBGraph
file: Given a $K2$, the index returns the chunk position in the
MRBGraph file.  As only point lookup is required, we employ a
hash-based implementation for the index.  The index is stored in an
index file and is preloaded into memory before Reduce computation.  We
apply the index nested loop join for the merging operation.

Can we further improve the join performance?  We observe that the
MapReduce shuffling phase will sort the intermediate keys.  As
seen in Section~\ref{subsec:engine}, the records in both the delta
MRBGraph and the MRBGraph file are in the order generated by the
shuffling phase.  That is, the two files are sorted in $K2$ order.
Therefore, we introduce a read cache and a dynamic read window
technique for further optimization.  Fig.~\ref{fig:mrbgstore} shows
the idea.  Given a sequence of $K2$s, there are two ways to read the
corresponding chunks:  (i) performing an individual I/O operation for
each chunk; or (ii) performing a large I/O that covers all the
required chunks.  The former may lead to frequent disk seeks, while
the latter may result in reading a lot of useless data.  Fortunately,
we know the list of sorted $K2$s to be queried.  Using the index, we
obtain their chunk positions.  We can estimate the costs of using a
large I/O vs. a number of individual I/Os, and intelligently determine
the read window size $w$ based on the cost estimation.

Algorithm~\ref{alg:query} shows the query algorithm to retrieve the
the chunk $k$ given a query key $k$ and the list of queried keys $L=\{L_1,L_2,\ldots\}$.  If the chunk $k$ does not reside in the read cache (line 1), it will compute the read window size $w$ by a heuristic, and read $w$ bytes into the read cache. The loop (line 4--8) probes the gap between two consecutive queried chunks (chunk $L_i$ and chunk $L_{i+1}$). The gap size indicates the wasted read effort. If the gap is less than a threshold $T$ ($T=100KB$ by default), we consider that the benefit of large I/O can compensate for the wasted read effort, and enlarge the window to cover chunk $L_{i+1}$. In this way, the algorithm finds the read window size $w$ by balancing the cost of a large I/O vs. a number of individual I/Os. It also ensures that the read window size does not exceed the read cache.  Then the algorithm read the next $w$ bytes into the read cache (line 9) and retrieves the requested chunk $k$ from the read cache (line 11).

\Paragraph{Incremental Storage of MRBGraph Changes} As shown in
Fig.~\ref{fig:mrbgstore}, the outputs of the merge operation, which
are the up-to-date MRBGraph states (chunks), are used to invoke the
Reduce function. In addition, the outputs are also buffered in an
append buffer in memory.  When the append buffer is full, the
MRBG-Store performs sequential I/Os to append the contents of the
buffer to the end of the MRBGraph file.  When the merge operation
completes, the MRBG-Store flushes the append buffer, and updates the
index to reflect the new file positions for the updated chunks.  Note
that obsolete chunks are NOT immediately updated in the file (or
removed from the file) for I/O efficiency.  The MRBGraph file is
reconstructed off-line when the worker is idle.  In this way, the
MRBG-Store efficiently supports incremental storage of MRBGraph
Changes.

As a result of the incremental storage, the MRBGraph file may contain
multiple segments of sorted chunks, each resulting from a merge
operation.  This situation frequently appears in iterative incremental
computation, for which we enhance the above query algorithm with a
multi-window technique to efficiently process the multiple segments.
We defer the in-depth discussion to Section~\ref{sec:incriter}.

\subsection{Optimization for Special Accumulator Reduce}
\label{subsec:accumulator}

We study a special case that appears frequently in applications and is
amenable to further optimization.  Specifically, the Reduce function
is an accumulative operation '$\oplus$':
\begin{equation}\label{eq:accumulate}
  f(\{V2_0, V2_1, ..., V2_k\})= V2_0 \oplus V2_1 \oplus\dots \oplus V2_k,\notag
\end{equation}
which satisfies the distributive property:
\begin{equation}\label{eq:accumulate2}
  f(D \cup \Delta D)=f(D)\oplus f(\Delta D),\notag
\end{equation}
and the incremental data set $\Delta D$ contains only insertions
without deletions or updates.  This property allows us to process the
two data set $D$ and $\Delta D$ separately and then to simply combine
the results by the '$\oplus$' operation to obtain the full result. We
call this kind of Reduce function \emph{accumulator Reduce}.
For this special case, it is not necessary to preserve the MRBGraph.
The engine will optimize the special case by only preserving the
Reduce output kv-pairs $\langle K3,V3\rangle$.  Then it simply invokes
the accumulator Reduce to accumulate changes to the result kv-pairs.

Many MapReduce algorithms employ accumulator Reduce. A well-known
example is WordCount. The Reduce function of WordCount computes the
count of word appearances using an integer sum operation, which
satisfies the above property.  Other common operations that directly
satisfy the distributive property include maximum and minimum.
Moreover, some operations can be easily modified to satisfy the
requirement of accumulator Reduce.  For example, average is computed
as dividing sum by count.  While it is not possible to combine two
averages into a single average, we can modify the implementation to
allow/produce a partial sum and a partial count in the function input
and the output.  Then the implementation can accumulate partial sums
and partial counts in order to compute the average of the full data
set.

To use this feature, a programmer should declare the accumulative
operation '$\oplus$' using a new interface \texttt{AccumulatorReducer}
in the MapReduce driver program (see Table \ref{tab:api}).

\section{General-Purpose Support for \\Iterative Computation}
\label{sec:iter}

We first analyze several representative iterative algorithms in
Section~\ref{sec:examples}.  Based on this analysis, we propose a
general-purpose MapReduce model for iterative computation in
Section~\ref{subsec:iterapi}, and describe how to efficiently support
this model in Section~\ref{subsec:iterimpl}.

\subsection{Analyzing Iterative Computation}
\label{sec:examples}

\Paragraph{PageRank} PageRank~\cite{Brin:1998:ALH:297810.297827} is a
well-known iterative graph algorithm for ranking web pages. It computes
a ranking score for each vertex in a graph.  After initializing all
ranking scores, the computation performs a MapReduce job per
iteration,  as shown in Algorithm~\ref{alg:pagerank}. $i$ and $j$ are
vertex ids, $N_i$ is the set of out-neighbor vertices of $i$, $R_i$ is
$i$'s ranking score that is updated iteratively.  `$|$' means
concatenation.  All $R_i$'s are initialized to one\footnote{The computed
PageRank scores will be $|N|$ times larger, where $|N|$ is the number of
vertices in the graph.}. The Map instance on vertex $i$ sends value
$R_{i,j}=R_i/|N_i|$ to all its out-neighbors $j$, where $|N_i|$ is the
number of $i$'s out-neighbors.  The Reduce instance on vertex $j$
updates $R_j$ by summing the $R_{i,j}$ received from all its
in-neighbors $i$, and applying a damping factor $d$.

\begin{table*}[t!]
\caption{Structure and state kv-pairs in representative iterative
algorithms.}
\label{tab:kvpair}
\centering
\small
\begin{tabular}{|c|c|c|c|c|c|}
\hline
 \textbf{Algorithm} & \textbf{Structure Key (SK)} & \textbf{Structure
Value (SV)} & \textbf{State Key (DK)} & \textbf{State Value (DV)} &
\textbf{SK $\leftrightarrow$ DK} \\
\hline
\hline
 PageRank & vertex id $i$ & out-neighbor set $N_i$ & vertex id $i$ & rank score $R_i$ & one-to-one \\
\hline
 Kmeans & point id $pid$ & point value $pval$ & unique key 1 & centroids $\{\langle cid,cval\rangle\}$ & all-to-one \\
\hline
 GIM-V & matrix block id $(i,j)$ & matrix block $m_{i,j}$ & vector block id $j$ & vector block $v_j$ & many-to-one \\
\hline
\end{tabular}
\end{table*}

\renewcommand{\algorithmicrequire}{\textbf{Map Phase}}
\renewcommand{\algorithmicensure}{\textbf{Reduce Phase}}

\begin{algorithm}
\caption{PageRank in MapReduce}
\label{alg:pagerank}
\begin{algorithmic}[1]
\REQUIRE input: $<i$, $N_i|R_i>$
\STATE output $<i$, $N_i>$
\FORALL{$j$ in $N_i$}
\STATE $R_{i,j}=\frac{R_i}{|N_i|}$
\STATE output $<j$, $R_{i,j}>$
\ENDFOR
\item[]
\ENSURE input: $<j$, $\{R_{i,j},N_j\}>$
\STATE $R_j=d\sum_i R_{i,j} + (1-d)$
\STATE output $<j$, $N_j | R_j>$
\end{algorithmic}
\end{algorithm}

\Paragraph{Kmeans} Kmeans~\cite{lloyd1982least} is a commonly used
clustering algorithm that partitions points into $k$ clusters.  We
denote the ID of a point as $pid$, and its feature values $pval$.  The
computation starts with selecting $k$ random points as cluster
centroids set $\{cid,cval\}$. As shown in Algorithm~\ref{alg:kmeans},
in each iteration, the Map instance on a point $pid$ assigns the point
to the nearest centroid.  The Reduce instance on a centroid $cid$
updates the centroid by averaging the values of all assigned points
$\{pval\}$.

\begin{algorithm}
\caption{Kmeans in MapReduce}
\label{alg:kmeans}
\begin{algorithmic}[1]
\REQUIRE input: $<pid$, $pval|\{cid,cval\}>$
\STATE $cid\leftarrow$ find the nearest centroid of $pval$ in
$\{cid,cval\}$
\STATE output $<cid$, $pval>$
\item[]
\ENSURE input: $<cid$, $\{pval\}>$
\STATE $cval\leftarrow$ compute the average of $\{pval\}$
\STATE output $<cid$, $cval>$
\end{algorithmic}
\end{algorithm}

\Paragraph{GIM-V} Generalized Iterated Matrix-Vector multiplication
(GIM-V)~\cite{5360248} is an abstraction of many iterative graph
mining operations (e.g., PageRank, spectral clustering, diameter
estimation, connected components).  These graph mining algorithms can
be generally represented by operating on an $n\times n$ matrix $M$ and
a vector $v$ of size $n$.  Suppose both the matrix and the vector are
divided into sub-blocks.  Let $m_{i,j}$ denote the $(i,j)$-th block of
$M$ and $v_j$ denote the $j$-th block of $v$.  The computation steps
are similar to those of the matrix-vector multiplication and can be
abstracted into three operations: (1) $mv_{i,j}$ =
\texttt{combine2}($m_{i,j},v_j$); (2) $v_i'$ =
\texttt{combineAll}$_i$(\{$mv_{i,j}$\}); and (3) $v_i$ =
\texttt{assign}($v_i,v_i'$).  We can compare \texttt{combine2} to the
multiplication between $m_{i,j}$ and $v_j$, and compare
\texttt{combineAll} to the sum of $mv_{i,j}$ for row $i$.
Algorithm~\ref{alg:gimv} shows the MapReduce implementation with two
jobs for each iteration. The first job assigns vector block $v_j$ to
multiple matrix blocks $m_{i,j}$ ($\forall i$) and performs
\texttt{combine2}($m_{i,j},v_j$) to obtain $mv_{i,j}$. The second job
groups the $mv_{i,j}$ and $v_i$ on the same $i$, performs the
\texttt{combineAll(\{$mv_{i,j}$\})} operation, and updates $v_i$ using
\texttt{assign}($v_i,v_i'$).

\begin{algorithm}
\caption{GIM-V in MapReduce}
\label{alg:gimv}
\begin{algorithmic}[1]
\REQUIRE1 input: $<(i,j), m_{i,j}>$ or $<j,v_j>$
\IF{kv-pair is $<(i,j), m_{i,j}>$}
\STATE output $<(i,j), m_{i,j}>$
\ELSIF{kv-pair is $<j,v_j>$}
\FORALL{$i$ blocks in $j$'s row}
\STATE output $<(i,j), v_j>$
\ENDFOR
\ENDIF
\item[]
\ENSURE1 input: $<(i,j), \{m_{i,j},v_j\}>$
\STATE $mv_{i,j}$ = combine2($m_{i,j}$, $v_j$)
\STATE output $<i$, $mv_{i,j}>$, $<j$, $v_j>$
\item[]
\REQUIRE2: output all inputs
\item[]
\ENSURE2 input: $<i, \{mv_{i,j},v_i\}>$
\STATE $v_i'\leftarrow$ combineAll(\{$mv_{i,j}$\})
\STATE $v_i\leftarrow$ assign($v_i$, $v_i'$)
\STATE output $<i$, $v_i>$
\end{algorithmic}
\end{algorithm}

\Paragraph{Two Kinds of Data Sets in Iterative Algorithms} From the
above examples, we see that iterative algorithms usually involve two
kinds of data sets: (i) loop-invariant \textit{structure data}, and
(ii) loop-variant \textit{state data}.  Structure data often reflects
the problem structure and is read-only during computation.  In
contrast, state data is the target results being updated in each
iteration by the algorithm.  Structure (state) data can be represented
by a set of \textit{structure (state) kv-pairs}. Table~\ref{tab:kvpair} displays the structure and state kv-pairs of
the three example algorithms.

\Paragraph{Dependency Types between State and Structure Data} There
are various types of dependencies between state and structure data, as
listed in Table~\ref{tab:kvpair}.  PageRank sees one-to-one
dependency: every vertex $i$ is associated with both an out-neighbor
set $N_i$ and a ranking score $R_i$.  In Kmeans, the Map instance of
every point requires the set of all centroids, showing an
all-to-one dependency.  In GIM-V, multiple matrix blocks $\forall j,
m_{i,j}$ are combined to compute the $i$th vector block $v_i$, thus the dependency is many-to-one.

\begin{figure}[t]
    \centering
        \includegraphics[width=3.2in]{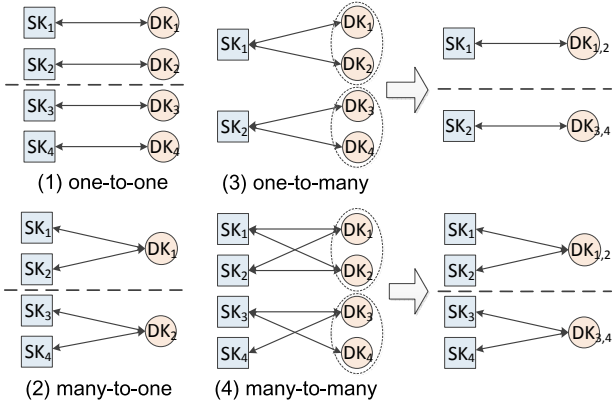}
    \caption{Dependency types between structure and state kv-pairs.
(3)/(4) can be converted into (1)/(2).}
    \label{fig:mapping}
\end{figure}

Generally speaking, there are four types of dependencies between
structure kv-pairs and state kv-pairs as shown in
Fig.~\ref{fig:mapping}: (1) one-to-one, (2) many-to-one, (3)
one-to-many, (4) many-to-many.  All-to-one (one-to-all) is a special
case of many-to-one (one-to-many).  PageRank is an example of (1).
Kmeans and GIM-V are examples of (2).  We have not encountered
applications with (3) or (4) dependencies.  (3) and (4) are listed
only for completeness of discussion.

In fact, for (3) one-to-many case and (4) many-to-many case, it is
possible to redefine the state key to convert them into (1) one-to-one
and (2) many-to-one dependencies, respectively, as show in the right
part of Fig.~\ref{fig:mapping}.  The idea is to re-organize the
MapReduce computation in an application or to define a custom
partition function for shuffling so that the state kv-pairs (e.g,
$DK_1$ and $DK_2$ in the figure) that Map to the same structure
kv-pair (e.g., $SK_1$ in the figure) are always processed in the same
task.  Then we can assign a key (e.g., $DK_{1,2}$) to each group of
state kv-pairs, and consider each group as a single state kv-pair.
Given this transformation, we need to focus on only (1) one-to-one and
(2) many-to-one cases.  Consequently, each structure kv-pair is interdependent with \textbf{ONLY} a single state kv-pair.   This is an
important property that we leverage in our design of i$^2$MapReduce.

\subsection{General-Purpose Iterative MapReduce Model}
\label{subsec:iterapi}

A number of recent efforts have been targeted at improving iterative
processing on MapReduce, including
Twister~\cite{Ekanayake:2010:TRI:1851476.1851593},
HaLoop~\cite{haloop}, and iMapReduce~\cite{imapreduce}.  In general,
the improvements focus on two aspects:

\begin{list}{\labelitemi}{\setlength{\leftmargin}{3mm}\setlength{\itemindent}{-1mm}\setlength{\topsep}{0mm}\setlength{\itemsep}{0mm}\setlength{\parsep}{1mm}}

\item \emph{Reducing job startup costs:} In vanilla MapReduce, every
algorithm iteration runs one or several MapReduce jobs.  Note that
Hadoop may take over 20 seconds to start a job with 10--100 tasks.  If
the computation of each iteration is relatively simple, job startup
costs may consist of an overly large fraction of the run time.  The
solution is to modify MapReduce to reuse the same jobs across
iterations, and kill them only when the computation completes.

\item \emph{Caching structure data:} Structure data is immutable
during computation.  It is also much larger than state data in many
applications (e.g., PageRank, Kmeans, and GIM-V).  Therefore, it is
wasteful to transfer structure data over and over again in every
iteration.  An optimization is to cache structure data in local file
systems to avoid the cost of network communication and reading from HDFS.

\end{list}

For the first aspect, we modify Hadoop to allow jobs to stay alive
across multiple iterations.

For the second aspect, however, a design must separate structure data
from state data, and consider how to match interdependent structure and
state data in the computation. HaLoop~\cite{haloop} uses an extra
MapReduce job to match structure and state data in each iteration.  We
would like to avoid such heavy-weight solution. iMapReduce~\cite{imapreduce} creates the same number of Map and Reduce
tasks, and connects every Reduce task to a Map task with a local connection to transfer the state data output from a Reduce task to the
corresponding Map task. However, this approach assumes one-to-one dependency for join operation.  Thus, it cannot support
Kmeans or GIM-V.

In the following, we propose a design that generalizes previous
solutions to efficiently support various dependency types.

\Paragraph{Separating Structure and State Data in Map API} We enhance
the Map function API to explicitly express structure vs. state
kv-pairs in i$^2$MapReduce:
\begin{equation}\label{eq:map}
  \texttt{map}(SK,SV,DK,DV)\rightarrow [\langle K2,V2\rangle]\notag
\end{equation}
\noindent The interdependent structure kv-pair $\langle SK,SV\rangle$ and
state kv-pair $\langle DK,DV\rangle$ are conjointly used in the
Map function.   A Map function outputs intermediate kv-pairs $\langle
K2,V2\rangle$s. The Reduce interface is kept the same as before.  A Reduce function
combines the intermediate kv-pairs $\langle K2,\{V2\}\rangle$s and
outputs $\langle K3,V3\rangle$:
\begin{equation}\label{eq:reduce}
  \texttt{reduce}(K2,\{V2\})\rightarrow \langle K3,V3\rangle\notag
\end{equation}

\begin{figure}[t]
    \centering
    \includegraphics[width=3.2in]{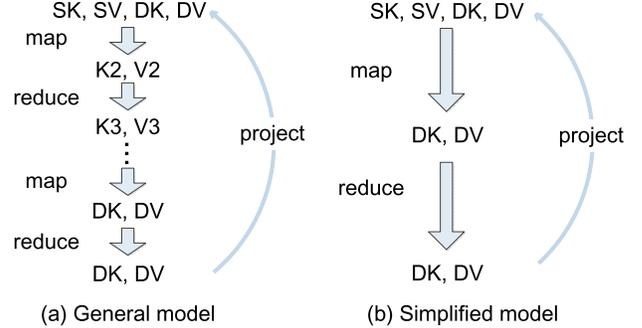}
    \caption{Iterative model of i$^2$MapReduce.}
    \label{fig:keyvalues}
\end{figure}

\Paragraph{Specifying Dependency with Project}
We propose a new API function, \emph{Project}.  It specifies the
interdependent state key of a structure key:
\begin{equation}\label{eq:project}
\begin{aligned}
  &\texttt{project}(SK)\rightarrow DK\notag
\end{aligned}
\end{equation}
Note that each structure kv-pair is interdependent with a single state
kv-pair.  Therefore, Project returns a single value $DK$ for each input $SK$.

\Paragraph{Iterative Model} Fig.~\ref{fig:keyvalues} shows our
iterative model.  By analyzing the three representative applications,
we find that the input of an iteration contains both structure and
state data, while the output is only the state data.  A large number
of iterative algorithms (e.g., PageRank and Kmeans) employs a single
MapReduce job in an iteration.  Their computation can be illustrated
using the simplified model as shown in Fig.~\ref{fig:keyvalues} (b).
In general, one or more MapReduce jobs may be used to update the
state kv-pairs $\langle DK,DV\rangle$, as shown in
Fig.~\ref{fig:keyvalues} (a).  Once the updated $\langle
DK,DV\rangle$s are obtained, they are matched to the interdependent
structure kv-pairs $\langle SK,SV\rangle$s with the Project function
for next iteration.  In this way, a kv-pair transformation loop is
built.  We call the first Map phase in an iteration the \emph{prime
Map} and the last Reduce phase in an iteration as the \emph{prime
Reduce}.

\subsection{Supporting Diverse Dependencies between Structure and State Data}
\label{subsec:iterimpl}

\Paragraph{Dependency-aware Data Partitioning} To support parallel
processing in MapReduce,  we need to partition the data.  Note that
both structure and state kv-pairs are required to invoke the Map
function.  Therefore, it is important to assign the interdependent structure kv-pair and
state kv-pair to the same partition so as to avoid
unnecessary network transfer overhead. Many existing systems such as Spark \cite{sparknsdi} and Stratosphere \cite{Ewen:2012:SFI:2350229.2350245} have applied this optimization. In i$^2$MapReduce, we design the following
partition function~(\ref{eq:partstate}) for state
and~(\ref{eq:partstructure}) for structure kv-pairs:
\begin{equation}\label{eq:partstate}
partition\_id=\texttt{hash}(DK, n) \hspace{0.7in}
\end{equation}
\begin{equation}\label{eq:partstructure}
partition\_id=\texttt{hash}(\texttt{project}(SK), n)
\end{equation}
where $n$ is the desired number of Map tasks.  Both functions employ
the same hash function.  Since Project returns the interdependent
$DK$ for a given $SK$, the interdependent $\langle SK,SV\rangle$s and
$\langle DK,DV\rangle$s will be assigned to the same partition.
i$^2$MapReduce partitions the structure data and
state data as the preprocessing step before an iterative job.


\Paragraph{Invoking Prime Map} i$^2$MapReduce launches a prime Map task
per data partition.  The structure and state kv-pairs assigned to a
partition are stored in two files: (i) a structure file containing
$\langle SK,SV\rangle$s and (ii) a state file containing $\langle
DK,DV\rangle$s.  The two files are provided as the input to the prime
Map task.  The state file is sorted in the order of $DK$, while the
structure file is sorted in the order of \texttt{project}($SK$).  That
is, the interdependent $SK$s and $DK$s are sorted in the same order.
Therefore, i$^2$MapReduce can sequentially read and match all the
interdependent structure/state kv-pairs through \emph{a single pass} of
the two files, while invoking the Map function for each matching pair.

\Paragraph{Task Scheduling: Co-locating Interdependent Prime Reduce and Prime Map} As shown
in Fig.~\ref{fig:keyvalues}, the prime Reduce computes the updated
state kv-pairs.  For the next iteration, i$^2$MapReduce must transfer
the updated state kv-pairs to their corresponding prime Map task, which caches their dependent structure kv-pairs in its local file
system.

The overhead of the backward transfer can be fully removed if the
number of state kv-pairs in the application is greater than or equal
to $n$, the number of Map tasks (e.g., PageRank and GIM-V).  The idea
is to create $n$ Reduce tasks, assign Reduce task $i$ to co-locate
with Map task $i$ on the same machine node, and make sure that Reduce
task $i$ produces and only produces the state kv-pairs in partition
$i$.  The latter can be achieved by employing the hash function of the
partition functions~(\ref{eq:partstate}) and~(\ref{eq:partstructure}) as
the shuffle function immediately before the prime Reduce phase.  The
Reduce output can be stored into an updated state file without any
network cost.  Interestingly, the state file is automatically sorted
in $DK$ order thanks to MapReduce's shuffle implementation.  In this
way, i$^2$MapReduce will be able to process the prime Map task of the
next iteration.

\Paragraph{Supporting Smaller Number of State kv-pairs}  In some
applications, the number of state keys is smaller than $n$.  Kmeans is
an extreme case with only a single state kv-pair.  In these
applications, the total size of the state data is typically quite
small.  Therefore, the backward transfer overhead is low.  Under such
situation, i$^2$MapReduce does not apply the above partition
functions.  Instead, it partitions the structure kv-pairs using
MapReduce's default approach, while replicating the state data to each
partition.

\section{Incremental Iterative Processing}
\label{sec:incriter}

In this section, we present incremental processing techniques for
iterative computation.  Note that it is not sufficient to simply
combine the above solutions for incremental one-step processing (in
Section~\ref{sec:incr}) and iterative computation (in
Section~\ref{sec:iter}).  In the following, we discuss three aspects
that we address in order to achieve an effective design.

\subsection{Running an Incremental Iterative Job}
\label{subsec:starti2}

Consider a sequence of jobs $A_1$, ... $A_i$, ... that
\emph{incrementally} refresh the results of an \emph{iterative}
algorithm.  Incoming new data and updates change the problem structure
(e.g., edge insertions or deletions in the web graph in PageRank, new
points in Kmeans, updated matrix data in GIM-V).  Therefore, structure
data evolves across subsequent jobs.  Inside a job, however, structure
data stays constant, but state data is iteratively updated and
converges to a fixed point.  The two types of data must be handled
differently when starting an incremental iterative job:

\begin{list}{\labelitemi}{\setlength{\leftmargin}{3mm}\setlength{\itemindent}{-1mm}\setlength{\topsep}{0mm}\setlength{\itemsep}{0mm}\setlength{\parsep}{1mm}}

\item \emph{Delta structure data:}  We partition the new data and
updates based on Equation~(\ref{eq:partstructure}), and
generate a delta structure input file per partition.

\item \emph{Previously converged state data:}  Which state shall we
use to start the computation?  For job $A_{i}$, we choose to use the
converged state data $D_{i-1}$ from job $A_{i-1}$, rather than the
random initial state $D_0$ (e.g., random centroids in Kmeans) for two
reasons.  First, $D_{i-1}$ is likely to be very similar to the
converged state $D_{i}$ to be computed by $A_{i}$ because there are
often only slight changes in the input data.  Hence, $A_i$ may
converge to $D_{i}$ much faster from $D_{i-1}$ than from $D_0$.
Second, only the states in the last iteration of $A_{i-1}$ need to be
saved.  If $D_0$ were used, the system would have to save the states
of every iteration in $A_{i-1}$ in order to incrementally process the
corresponding iteration in $A_{i}$.  Thus, our choice can
significantly speed up convergence, and reduce the time and space
overhead for saving states.

\end{list}


To run an incremental iterative job $A_i$, i$^2$MapReduce treats each
iteration as an incremental one-step job as shown previously in
Fig.~\ref{fig:increxam}.  In the first iteration, the delta input is
the delta structure data.  The preserved MRBGraph reflects the last
iteration in job $A_{i-1}$.  Only the Map and Reduce instances that
are affected by the delta input are re-computed.  The output of the
prime Reduce is the delta state data.  Apart from the computation,
i$^2$MapReduce refreshes the MRBGraph with the newly computed
intermediate states.  We denote the resulting updated MRBGraph as
MRBGraph$_1$.

In the $j$-th iteration ($j \geq 2$), the structure data remains the
same as in the $(j-1)$-th iteration, but the loop-variant state data
have been updated.  Therefore, the delta input is now the delta state
data.  Using the preserved MRBGraph$_{j-1}$, i$^2$MapReduce
re-computes only the Map and Reduce instances that are affected by the
input change.  It preserves the newly computed intermediate states in
MRBGraph$_j$.  It computes a new delta state data for the next
iteration.

The job completes when the state data converges or certain predefined
criteria are met.  At this moment, i$^2$MapReduce saves the converged
state data to prepare for the next job $A_{i+1}$.

\subsection{Extending MRBG-Store for Multiple Iterations}
\label{sec:emrbgstore}

As described previously in Section~\ref{subsec:mrbgstore}, MRBG-Store
appends newly computed chunks to the end of the MRBGraph file and
updates the chunk index to reflect the new positions.  Obsolete chunks
are removed offline when the worker machine is idle.
In an incremental iterative job, every iteration will generate newly
computed chunks, which are sorted due to the MapReduce shuffling
phase.  Consequently, the MRBGraph file will consist of multiple
batches of sorted chunks, corresponding to a series of iterations.  If
a chunk exists in multiple batches, a retrieval request returns the
latest version of the chunk (as pointed to by the chunk index).  In
the following, we extend the query algorithm
(Algorithm~\ref{alg:query}) to handle multiple batches of sorted
chunks.

\begin{figure}[t]
    \centering
        \includegraphics[width=3.1in]{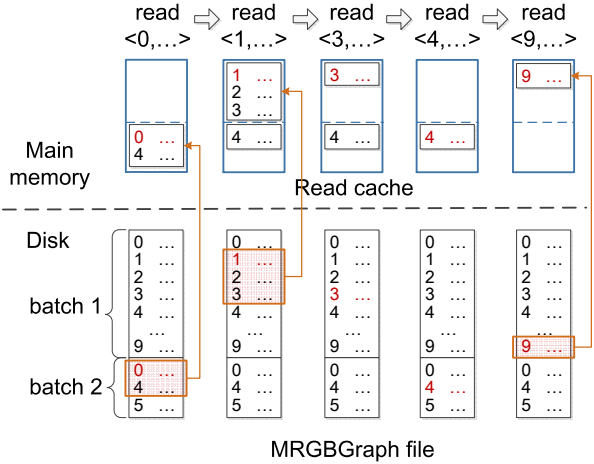}
    \caption{An example of reading a sequence of chunks with key
0,1,3,4,9,\ldots by using multi-dynamic-window.}
    \label{fig:emrbgstore}
\end{figure}

We propose a \textit{multi-dynamic-window} technique. Multiple dynamic
windows correspond to multiple batches (iterations).
Fig.~\ref{fig:emrbgstore} illustrates how the multi-dynamic-window
technique works via an example.  In this example, the MRBGraph file
contains two batches of sorted chunks.  It is queried to retrieve five
chunks as shown from left to right in the figure.  Note that the chunk
retrieval requests are sorted because of MapReduce's shuffling
operation.
The algorithm creates two read windows, each in charge of reading
chunks from the associated batch.  Since the chunks are sorted, a read
window will only slide downward in the figure.  The first request is
for chunk 0.  It is a read cache miss.  Although chunk 0 exists in
both batches, the chunk index points to the latest version in batch 2.
At this moment, we apply the analysis of Line 4--8 in
Algorithm~\ref{alg:query}, which determines the size of the I/O read
window.  The only difference is that we skip chunks that do not reside
in the current batch (batch 2).  As shown in
Fig.~\ref{fig:emrbgstore}, we find that it is profitable to use a
larger read window so that chunk 4 can also be retrieved into the read
cache.  The request for chunk 1 is processed similarly.  Chunk 0 is
evicted from the read cache because retrieval requests are always
non-decreasing.  The next two requests are for chunk 3 and chunk 4.
Fortunately, both of the chunks have been retrieved along with
previous requests.  The two requests hit in the read cache.  Finally,
the last request is satisfied by reading chunk 9 from batch 1.  Since
there are no further requests, we use the smallest possible read
window in the I/O read.

Even though MRBG-Store is designed to optimize I/O performance, the MRBGraph maintenance could still result in significant I/O cost. The I/O cost might outweigh the savings of incremental processing. For example, for applications with accumulator Reduce, MRBGraph is not
necessary for incremental Reduce computation, and therefore it is
advisable to turn off MRBGraph maintenance. Moreover, for Kmeans
computation, a single state value contains all the centroids.
Therefore, any updates in the input data will result in the change in
the state data, which will lead to global re-computation in the
subsequent iterations.  In this case, maintaining MRBGraph is
wasteful. It is better to only use iterative processing engine without
using MRBGraph. By analyzing the iterative computation's property, users have the option to turn on or turn off the MRBGraph maintenance functionality.

For incremental processing, i$^2$MapReduce maintains MRBGraph by default. However, the framework is able to detect the over-costly situation and automatically turn off MRBGraph maintenance. Consider an sequence of iterative computations $1,2,\ldots,i-1$ that converge at iteration $i-1$. The converged state $D_{i-1}$ and the converged intermediate computation state MRBGraph$_{i-1}$ are preserved for future usage. Recall that (Section \ref{subsec:starti2}), as the structure data are changed (reflected in the delta input), we start incremental iteration $i$ by using $D_{i-1}$ and MRBGraph$_{i-1}$. Only the Map and Reduce instances that are affected by the delta input are re-computed. The output of the prime Reduce is the \textbf{delta} state data $\Delta D_{i}$, which is a part of the whole updated state data $D_{i}$. In the next iteration, the delta input becomes the delta state data $\Delta D_{i}$. i$^2$MapReduce re-computes the Map and Reduce instances that are affected by $\Delta D_{i}$. Therefore, the proportion of the delta state data size to the entire state data size, i.e., $P_{\Delta}=\frac{|\Delta D_{i}|}{|D_{i-1}\cup\Delta D_{i}|}$ implies the amount of recomputations. The larger $P_{\Delta}$ the more recomputations. i$^2$MapReduce detects the size proportion $P_{\Delta}$ and turns off MRBGraph maintenance when $P_{\Delta}$ is larger than a threshold (50\% by default). For example, the Kmeans computation leads to $P_{\Delta}=100\%$. The framework will turn off MRBGraph maintenance and perform computation with only iterative processing support.

\subsection{Reducing Change Propagation}
\label{subsec:cpc}

In incremental iterative computation, changes in the delta input may
propagate to more and more kv-pairs as the computation iterates.  For
example, in PageRank, a change that affects a vertex in a web graph
propagates to the neighbor vertices after an iteration, to the
neighbors of the neighbors after two iterations, to the three-hop
neighbors after three iterations, and so on.  Due to this effect,
incremental processing may become less effective after a number of
iterations.

To address this problem, i$^2$MapReduce employs a \emph{change
propagation control} technique, which is similar to the \emph{dynamic computation} in GraphLab \cite{distgraphlab}.  It filters negligible changes of
state kv-pairs that are below a given threshold.   These filtered
kv-pairs are supposed to be very close to convergence.  Only the state
values that see changes greater than the threshold are emitted for
next iteration.  The changes for a state kv-pair are accumulated.  It is
possible a filtered kv-pair may later be emitted if its accumulated
change is big enough.

The observation behind this technique is that iterative
computation often converges asymmetrically:  Many state kv-pairs
quickly converge in a few iterations, while the remaining state
kv-pairs converge slowly over many iterations.  Low et al. has shown
that in PageRank computation the majority of vertices require only a
single update while only about 3\% of vertices take over 10 iterations
to converge~\cite{distgraphlab}.  Our previous work~\cite{priter} has
also exploited this property to give preference to the slowly
converged data items.

While this technique might impact result accuracy, the impact is
often minor since all ``influential'' kv-pairs would be
above the threshold and thus emitted.
This is indeed confirmed in our experiments
in Section~\ref{sec:cpropc}. If an application has high accuracy
requirement, the application programmer has the option to disable the
change propagation control functionality.

\section{Fault Tolerance and Load Balancing}
\label{sec:ftlb}

\subsection{Fault Tolerance}
\label{sec:ftlb:ft}

Vanilla MapReduce reschedules the failed Map/Reduce task in case task
failure is detected. However, the interdependency of prime Reduce
tasks and prime Map tasks in i$^2$MapReduce requires more complicated
fault-tolerance solution. i$^2$MapReduce checkpoints the prime Reduce
task's output state data and MRBGraph file on HDFS in every iteration.

Upon detecting a failure, i$^2$MapReduce recovers by considering task
dependencies in three cases.  (i) In case a prime Map task fails, the
master reschedules the Map task on the worker where its dependent
Reduce task resides.  The prime Map task reloads the its structure
data and resumes computation from its dependent state data
(checkpoint). (ii) In case a prime Reduce task fails, the master
reschedules the Reduce task on the worker where its dependent Map task
resides. The prime Reduce task reloads its MRBGraph file (checkpoint)
and resumes computation by re-collecting Map outputs. (iii) In case a
worker fails, the master reschedules the interdependent prime Map task
and prime Reduce task to a healthy worker together. The prime Map task
and Reduce task resume computation based on the checkpointed state
data and MRBGraph file as introduced above.

Following the design, we implement the fault tolerance mechanism. The failure recovery exploits the interdependency between prime Map tasks and prime Reduce tasks. The task scheduler on the master maintains the interdependency and the task-to-tracker\footnote{In Hadoop, TaskTracker is a process running on each slave node. It is in charge of executing each assigned Map/Reduce task.} assignment in a hash table. A task failure will be detected first by the TaskTracker, who will notify the master via heartbeat message (every 3 seconds by default). Upon receiving a task failure notification, the task scheduler on the master node looks up the task-to-tracker hash table and reassigns the failed task on the same TaskTracker. In case a worker (TaskTracker) fails, the task scheduler reassigns the interdependent prime Map task and prime Reduce task to another healthy worker. The reassigned task along with its current iteration information will be re-launched using the checkpointed data. The prime Map task reloads its structure data and resumes computation from its dependent state data (checkpoint). The prime Reduce task reloads its MRBGraph file (checkpoint) and resumes computation by re-collecting Map outputs.

\subsection{Load Balancing}
\label{sec:ftlb:lb}

Skewed structure data can lead to skewed workloads across workers.  To
deal with this problem, we can integrate online skew migration
technique \cite{Kwon:2012:SMS:2213836.2213840} to balance the
workload. Basically, it first identifies the task with the greatest
expected remaining processing time through probing. The unprocessed
input data of this straggling task is then proactively repartitioned
in a way that fully utilizes the nodes in the cluster and preserves
the ordering of the input data so that the original output can be
reconstructed by concatenation. In order to integrate online skew
migration into i$^2$MapReduce, the key challenge is to split and move
the task state (i.e., MRBGraph file) in an efficient way. The load
balancing mechanism is out of the scope of this paper and will be left
for future work.

\section{API Changes to MapReduce}
\label{sec:api}

\newcommand{\tabincell}[2]{\begin{tabular}{@{}#1@{}}#2\end{tabular}}
\begin{table*}[t]
\footnotesize
    \caption{API changes to Hadoop MapReduce}
    \label{tab:api}
    \centering
    \begin{tabular}{|l|l|l|l|}
    \hline
    \bfseries Job Type & \bfseries Functionality & \bfseries Vanilla MapReduce (Hadoop) & \bfseries i$^2$MapReduce \\
\hline\hline
Incremental One-Step & input format  & input: $\langle K1,V1'\rangle$ & delta input: $\langle K1,V1,`+'/`-'\rangle$ \\
\hline
Accumulator Reduce   & Reducer class   & \tabincell{l}{\texttt{reduce($K2$,\{$V2$\})$\rightarrow \langle K3,V3\rangle$}}
                     & \tabincell{l}{\texttt{accumulate($V2_{old},V2_{new}$)}$\rightarrow V2$}\\
\hline
\multirow{5}{*}{Iterative} & input format  & mixed input: $\langle K1,V1'\rangle$ & \tabincell{l}{structure input: $\langle SK,SV\rangle$\\state input: $\langle DK,DV\rangle$}\\\cline{2-4}
                            & Projector class &  & \tabincell{l}{\texttt{project($SK$)$\rightarrow DK$} \\\texttt{setProjectType(ONE2ONE)}} \\\cline{2-4}
                            & Mappper class & \tabincell{l}{\texttt{map($K1$, $V1$)$\rightarrow[\langle K2,V2\rangle]$}}
                                        & \tabincell{l}{\texttt{map}($SK$,$SV$,$DK$,$DV$)$\rightarrow [\langle K2,V2\rangle]$\\\texttt{init($DK$)$\rightarrow DV$}}\\
\hline
\multirow{3}{*}{Incremental Iterative} & input format & input: $\langle K1,V1'\rangle$ & delta structure input: $\langle SK,SV,`+'/`-'\rangle$\\\cline{2-4}
                                        & \tabincell{l}{change propagation\\control} &  & \tabincell{l}{\texttt{job.setFilterThresh(thresh)}\\\texttt{difference($DV_{curr}$,$DV_{prev}$)$\rightarrow diff$}}\\
\hline
\end{tabular}
\end{table*}

We implement a prototype of i$^2$MapReduce by modifying Hadoop-1.0.3. In order to support incremental and iterative processing, a few MapReduce APIs are changed or added. We summarize these API changes in Table \ref{tab:api}. We briefly explain the key APIs and their usage in this section.

\begin{list}{\labelitemi}{\setlength{\leftmargin}{3mm}\setlength{\itemindent}{-1mm}\setlength{\topsep}{0mm}\setlength{\itemsep}{0mm}\setlength{\parsep}{1mm}}

\item \emph{For incremental one-step processing}, programmers need to
specify the delta input, in which the inserted and deleted input
kv-pairs are marked with `$+$' and `$-$', respectively.

\item \emph{For the special case of accumulator reduce}, an
\texttt{accumulate} function needs to be specified, which aggregates
reducer input values with the same key.

\item \emph{For iterative computation}, programmers must specify the
structure kv-pairs $\langle SK,SV\rangle$, the state kv-pairs $\langle
DK,DV\rangle$, and the Project function.  Besides, a new mapper
interface should be implemented, and the new map
function will take both the structure and state kv-pairs as input.
The initial state value $DV$ should also be set.

\item \emph{For the incremental iterative computation}, in addition to
specifying the delta structure input, programmers can turn on the change
propagation control mechanism by setting the filter threshold and
specifying how to compute the change of a kv-pair given the current
and previous result values ($DV_{curr}$ and $DV_{prev}$).  Code
examples of various algorithms can be found on the project homepage\footnote{http://code.google.com/p/incr-iter-hadoop/}.

\end{list}

\section{Experiments}
\label{sec:expr}

In this section, we perform real-machine experiments to evaluate i$^2$MapReduce.

\subsection{Experiment Setup}

\subsubsection{Solutions to Compare} Our experiments compare four
solutions: (i) \emph{PlainMR recomp}, re-computation on vanilla Hadoop; (ii)
\emph{iterMR recomp}, re-computation on Hadoop optimized for iterative
computation (as described in Section~\ref{sec:iter}); (iii)
\emph{HaLoop recomp}, re-computation on the iterative MapReduce
framework HaLoop \cite{haloop}, which optimizes MapReduce by providing a structure
data caching mechanism; (iv) \emph{i$^2$MapReduce}, our proposed
solution.  To the best of our knowledge, the task-level coarse-grain
incremental processing system, Incoop \cite{incoop}, is not publicly available.
Therefore, we cannot compare i$^2$MapReduce with Incoop. Nevertheless,
our statistics show that without careful data partition, almost all tasks see changes in the
experiments, making task-level incremental processing less effective.

\subsubsection{Experimental Environment} All experiments run on Amazon
EC2.  We use 32 m1.medium instances. Each m1.medium instance is
equipped with 2 ECUs, 3.7GB memory, and 410GB storage.

\subsubsection{Applications} We have implemented four iterative mining
algorithms, including PageRank (one-to-one correlation), Single Source Shortest Path (SSSP, one-to-one correlation), Kmeans
(all-to-one correlation), and GIM-V (many-to-one correlation).  For
GIM-V, we implement iterative matrix-vector multiplication as the
concrete application using GIM-V model.

We also implemented a one-step mining algorithm, APriori
\cite{Agrawal:1994:FAM:645920.672836}, for mining frequent item sets. The APriori algorithm is used to compute the occurrence counts of frequent word pairs of a
Twitter data set.  After generating the candidate list of frequent
word pairs in a preprocessing job,  APriori runs a MapReduce job to
count the frequency of each word pair.  The Map task loads this list
into memory, and initializes a local count per pair.  Then, each input
tweet is processed by the Map function to identify any candidate pairs
and accumulate the associated local counts.  After this, the Map task
sends $\langle$word pair, local count$\rangle$ as intermediate
kv-pairs. Finally, the Reduce task aggregates the local counts into
the global frequency for each pair. Note that Apriori satisfies the requirements in Section~\ref{subsec:accumulator}.  Hence, we employ
the accumulator Reduce optimization in incremental processing.

\makesavenoteenv{tabular}
\makesavenoteenv{table}
\begin{table}[!t]
    \caption{Data sets}
    \label{tab:dataset}
    \centering
    \footnotesize
    \begin{tabular}{c|c|c|c}
    \hline
    algorithm & data set & size & description \\
\hline\hline
APriori & Twitter & 122 GB & 52,233,372 tweets\\
\hline
PageRank & ClueWeb & 36.4 GB & \tabincell{c}{20,000,000 pages\\365,684,186 links}\\
\hline
SSSP & ClueWeb2 & 70.2 GB & \tabincell{c}{20,000,000 pages\\365,684,186 links}\\
\hline
Kmeans & BigCross & 14.4 GB & \tabincell{c}{46,481,200 points\\57 dimensions} \\
\hline
GIM-V & WikiTalk & 5.4 GB & \tabincell{c}{100,000 rows\\1,349,584 non-0 entries} \\
\hline
\end{tabular}
\end{table}

\subsubsection{Data Sets}
Table~\ref{tab:dataset} describes the data sets for the five
applications.

The \textbf{Twitter} dataset data set is crawled from Aug. 1, 2011 to Sep. 30, 2011. It contains 52,233,372 tweets in JSON format and the size is about 122 GB. The APriori algorithm is performed to mine the frequent word pairs of the tweets.

The \textbf{ClueWeb} data set is a semi-synthetic data set generated from a base real-world data set \footnote{http://lemurproject.org/clueweb09/}. The original data set consists of 1,040,809,705 nodes (web pages) and 7,944,351,835 links, and its size is 71GB. Due to the high complexity resulted from the large number of nodes and links, we cannot complete the PageRank computation in a reasonable time period. Thus, we extracted 20,000,000 nodes and their 365,684,186 links from the original data set to form a smaller graph (6 GB). Further, we substituted all node identifiers with longer strings to make the structure data larger without changing the graph structure. The extended ClueWeb data set is 36.4 GB.

The \textbf{ClueWeb2} data set is generated from the ClueWeb data set. Since SSSP application runs on a weighted graph, we modify the ClueWeb graph by adding each edge with a random weight following gaussian distribution. Finally, the resulted ClueWeb2 data set is 70.2 GB.

The \textbf{BigCross} data set is a semi-synthetic data set generated from a high-volume and high-dimensional real data set \footnote{http://www.cs.uni-paderborn.de/en/fachgebiete/ag-bloemer/research/clustering/streamkmpp} . The original data set consists of 11,620,300 individuals and each is with 57 dimensions, the total size of which is 3.6 GB. We generate the BigCross data set by repeating the original data set four times to make it larger, so the size is 14.4 GB. We randomly pick 64 points from the whole data set as 64 initial centers.

The \textbf{WikiTalk} data set is also a semi-synthetic data set generated from a real world WikiTalk network data set \footnote{http://snap.stanford.edu/data/wiki-Talk.html}. The original WikiTalk network contains all the users and discussion from the inception of Wikipedia till January 2008. Nodes in the network represent Wikipedia users and a directed edge from node $i$ to node $j$ represents that user i at least once edited a talk page of user $j$. Therefore, we can generate a matrix data set based on the real world data, which is used in GIM-V (matrix-vector) computation. The original data set consists of 2,394,385 rows and 5,021,410 non-zero entries, and the size is 66.5 MB. Due to the high complexity of matrix-vector computation, we extracted 100,000 nodes and 1,349,584 non-zero entries from the original data set to form a smaller matrix. We also substituted all point identifiers with longer strings to make the data set larger. The extended WikiTalk data set is 5.4 GB.

\subsubsection{Delta Input, and Converged States} For incremental processing, we generate a delta input
from each data set.  For APriori, the Twitter dataset is collected over a period of two months.
We choose the last week's messages as the delta input, which is
7.9\% of the input.  For the four iterative algorithms, the delta
input is generated by randomly changing 10\% of the input data unless
otherwise noted.  To make the comparison as fair as possible, we start
incremental iterative processing \emph{from the previously converged states
for all the four solutions}.

\subsection{Overall Performance}

\Paragraph{Incremental One-Step Processing}  We use APriori to
understand the benefit of incremental one-step processing in
i$^2$MapReduce.  MapReduce re-computation takes 1608 seconds. In
contrast, i$^2$MapReduce takes only 131 seconds.  Fine-grain
incremental processing leads to a 12x speedup.

\begin{figure}[t]
    \centering
        \includegraphics[width=3in]{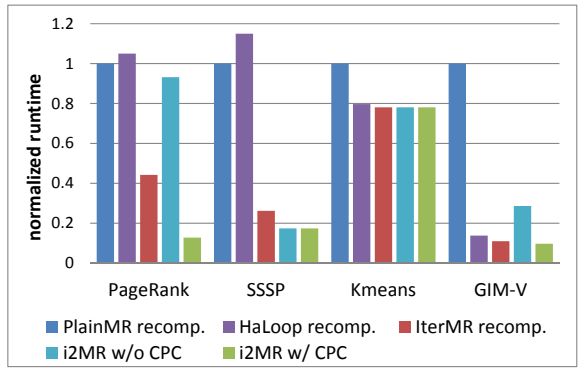}
    \caption{Normalized runtime.}
    \label{fig:runtime}
\end{figure}

\Paragraph{Incremental Iterative Processing} Fig. \ref{fig:runtime}
shows the normalized runtime of the four iterative algorithms while
10\% of input data has been changed.  ``1'' corresponds to the runtime
of PlainMR recomp.

For PageRank, iterMR reduces the runtime of PlainMR recomp by 56\%.
The main saving comes from the caching of structure data and the
saving of the MapReduce startup costs.  i$^2$MapReduce improves the
performance further with fine-grain incremental processing and change propagation control (CPC), achieving
a speedup of 8 folds (i2MR w/o CPC). We also show that without change propagation control the changes it will return the exact updated result but at the same time prolong the runtime (i2MR w/o CPC). The change propagation control technique is critical to guarantee the performance. Section \ref{sec:cpropc} will discuss the effect of CPC in more details. On the other hand, it is surprising to see
that HaLoop performs worse than plain MapReduce.  This is because
HaLoop employs an extra MapReduce job in each iteration to join the
structure and state data \cite{haloop}. The profit of caching cannot compensate for the extra cost when the structure data is not big enough. Note that the iterative model in
i$^2$MapReduce avoids this overhead by exploiting the Project function
to co-partition structure and state data. The detail comparison with HaLoop is provided in Section \ref{sec:vshaloop}.

For SSSP, the performance gain of i$^2$MapReduce is similar to that for PageRank. We set the filter threshold to 0 in the change propagation control. That is, nodes without any changes will be filtered out. Therefore, unlike PageRank, the SSSP results with CPC are precise.

For Kmeans, small portion of changes in input will lead to global
re-computation.  Therefore, we turn off the MRBGraph functionality. As
a result, i$^2$MapReduce falls back to iterMR recomp.  We see that
HaLoop and iterMR exhibit similar performance.  They both outperform
plainMR because of similar optimizations, such as caching structure
data.

For GIM-V, both plainMR and HaLoop run two MapReduce jobs in each
iteration, one of which joins the structure data (i.e., matrix) and the
state data (i.e., vector).  In contrast, our general-purpose iterative
support removes the need for this extra job.  iterMR and
i$^2$MapReduce see dramatic performance improvements.  i$^2$MapReduce
achieves a 10.3x speedup over plainMR, and a 1.4x speedup over HaLoop.

\subsection{Time Breakdown Into MapReduce Stages}

\begin{figure}[t]
    \centering
        \includegraphics[width=2.9in]{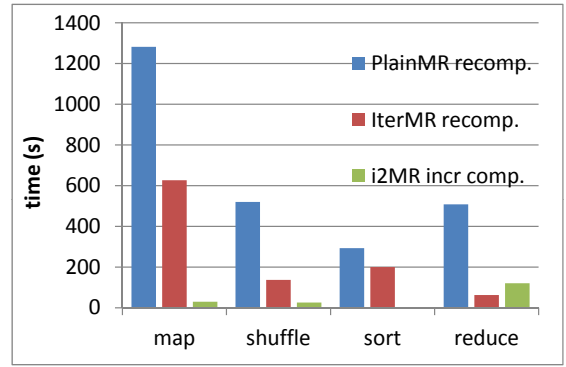}
    \caption{Run time of individual stages in PageRank.}
    \label{fig:phase}
\end{figure}

To better understand the overall performance, we report the
time\footnote{The resulted time does not include the structure data
partition time, while both the iterMR time and i2MR time in Fig.
\ref{fig:runtime} include the time of structure data partition job for
fairness.} of the individual MapReduce stages (across all iterations)
for PageRank in Fig. \ref{fig:phase}.

For the Map stage, IterMR improves the run time by 51\% because it
separates the structure and state data, and avoids reading and parsing
the structure data in every iteration.  i$^2$MapReduce further
improves the performance with fine-grain incremental processing,
reducing the plainMR time by 98\%.  Moreover, we find that the change
propagation control mechanism plays a significant role.  It filters
the kv-pairs with tiny changes at the prime Reduce, greatly decreasing
the number of Map instances in the next iteration.  (cf.
Section~\ref{sec:cpropc})

For the shuffle stage, iterMR reduces the run time of PlainMR by 74\%.
Most savings result from avoiding shuffling structure data from Map
tasks to Reduce tasks.  Moreover, compared to iterMR, i$^2$MapReduce
shuffles only the intermediate kv-pairs from the Map instances that
are affected by input changes, thereby further improving the shuffle
time, achieving 95\% reduction of PlainMR time.

For the sort stage, i$^2$MapReduce sorts only the small number of
kv-pairs from the changed Map instances, thus removing almost all
sorting cost of PlainMR.

For the Reduce stage, iterMR cuts the run time of PlainMR by 88\%
because it does not need to join the updated state data and the
structure data.  Interestingly, i$^2$MapReduce takes longer than
iterMR.  This is because i$^2$MapReduce pays additional cost for
accessing and updating the MRBGraph file in the MRBG-Store.  We study
the performance of MRBG-Store in the next subsection.

\subsection{Performance Optimizations in MRBG-Store}

As shown in Table~\ref{tab:mrbgstore},  we enable the optimization
techniques in MRBG-Store one by one for PageRank, and report three
columns of results: (i) total number of I/O reads in
Algorithm~\ref{alg:query} (which likely incur disk seeks), (ii) total
number of bytes read in Algorithm~\ref{alg:query}, and (iii) total
elapsed time of the merge operation.  (i) and (ii) are across all the
workers and iterations, and (iii) is across all the iterations.  Note
that the MRBGraph file maintains the intermediate data distributively,
the total size of which is 572.4 GB in the experiment.

\begin{table}[t]
    \caption{Performance optimizations in MRBG-Store}
    \label{tab:mrbgstore}
    \centering
    \footnotesize
    \begin{tabular}{|l|r|r|r|}
    \hline
    technique & \# reads & rsize(GB) & time (s) \\
\hline\hline
index-only & 5519910 & 34.2 & 718\\
\hline
single-fix-window & 1263680 & 10512.6 & 1361\\
\hline
multi-fix-window & 1188420 & 337.8 & 513 \\
\hline
multi-dynamic-window & 2418809 & 153.6 & 467 \\
\hline
\end{tabular}
\end{table}

First, only the chunk index is enabled.  For a given key, MRBG-Store
looks it up in the index to obtain the exact position of its chunk,
and then issues an I/O request to read the chunk.  This approach reads
only the necessary bytes but issues a read for each chunk.  As shown
in Table~\ref{tab:mrbgstore}, index-only has the smallest read size (rsize),
but incurs the largest number of I/O reads.

Second, with a single fix-sized read window,  a single I/O read may
cover multiple chunks that need to be merged, thus significantly
saving disk seeks.  However, since PageRank is an iterative algorithm
and multiple sorted batches of chunks exist in the MRBGraph file (cf.
Section \ref{sec:emrbgstore}), the next to-be-accessed chunk might not
reside in the same batch.  Consequently, this approach often wastes
time reading a lot of obsolete chunks.  Its elapsed time gets worse.

Third, we use multiple fix-sized windows for iterative computation.
This approach addresses the weakness of the single fix-sized window.
As shown in Table~\ref{tab:mrbgstore}, it dramatically reduces the
number of I/O reads and the bytes read from disks, achieving an 1.4x
improvement over the index-only case.

Finally, our solution in i$^2$MapReduce optimizes further by
considering the positions of the next chunks to be accessed and making
intelligent decisions on the read window sizes.  As a result,
multi-dynamic-window reads smaller amount of data.  It achieves a 1.6x
speedup over the index-only case.

\subsection{Effect of Change Propagation Control}
\label{sec:cpropc}

\begin{figure}[!t]
	\centerline{\subfloat[Runtime]{\includegraphics[width=1.25in, angle=-90]{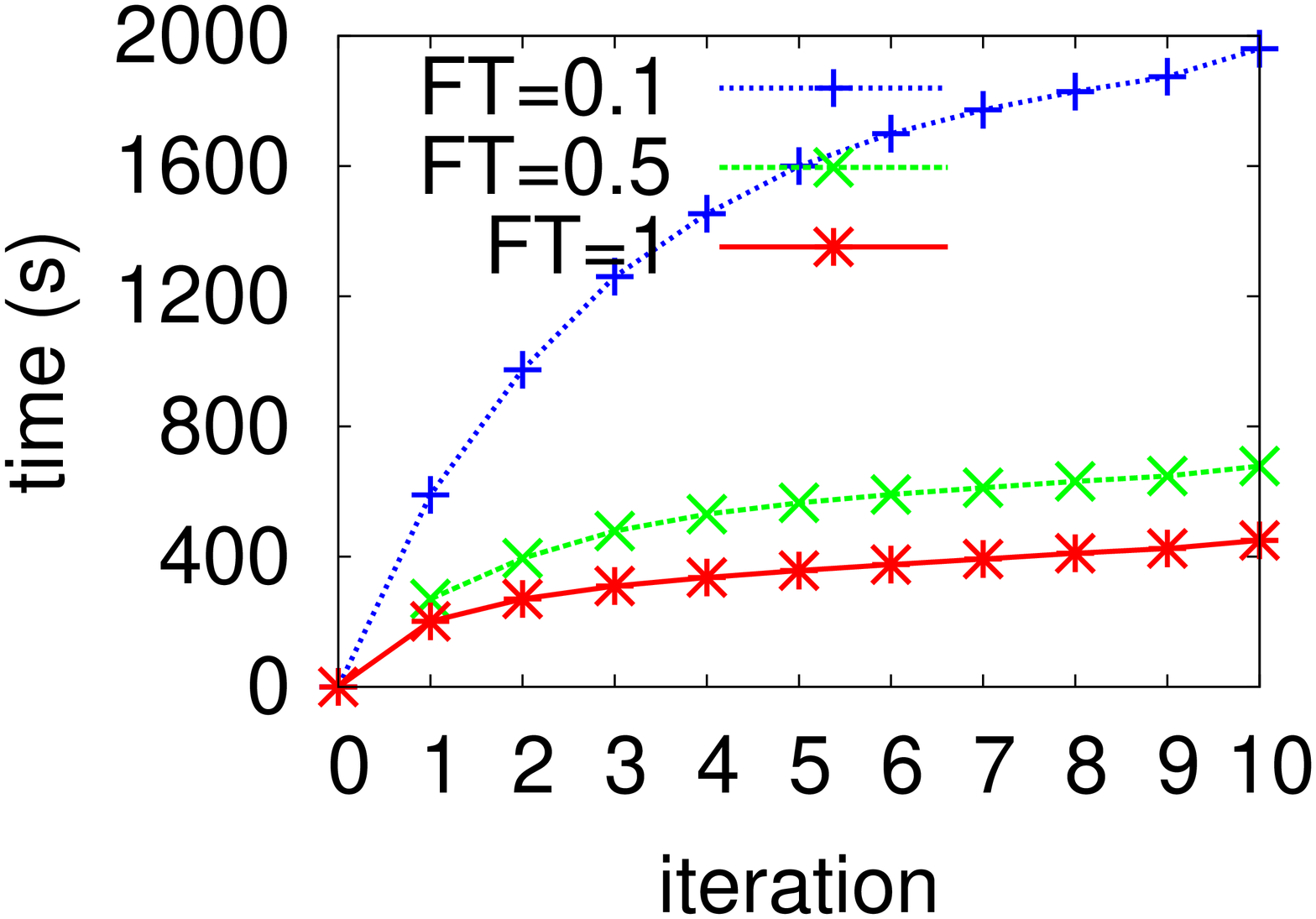}
    \label{fig:filter:time}}
	\subfloat[Mean error]{\includegraphics[width=1.25in, angle=-90]{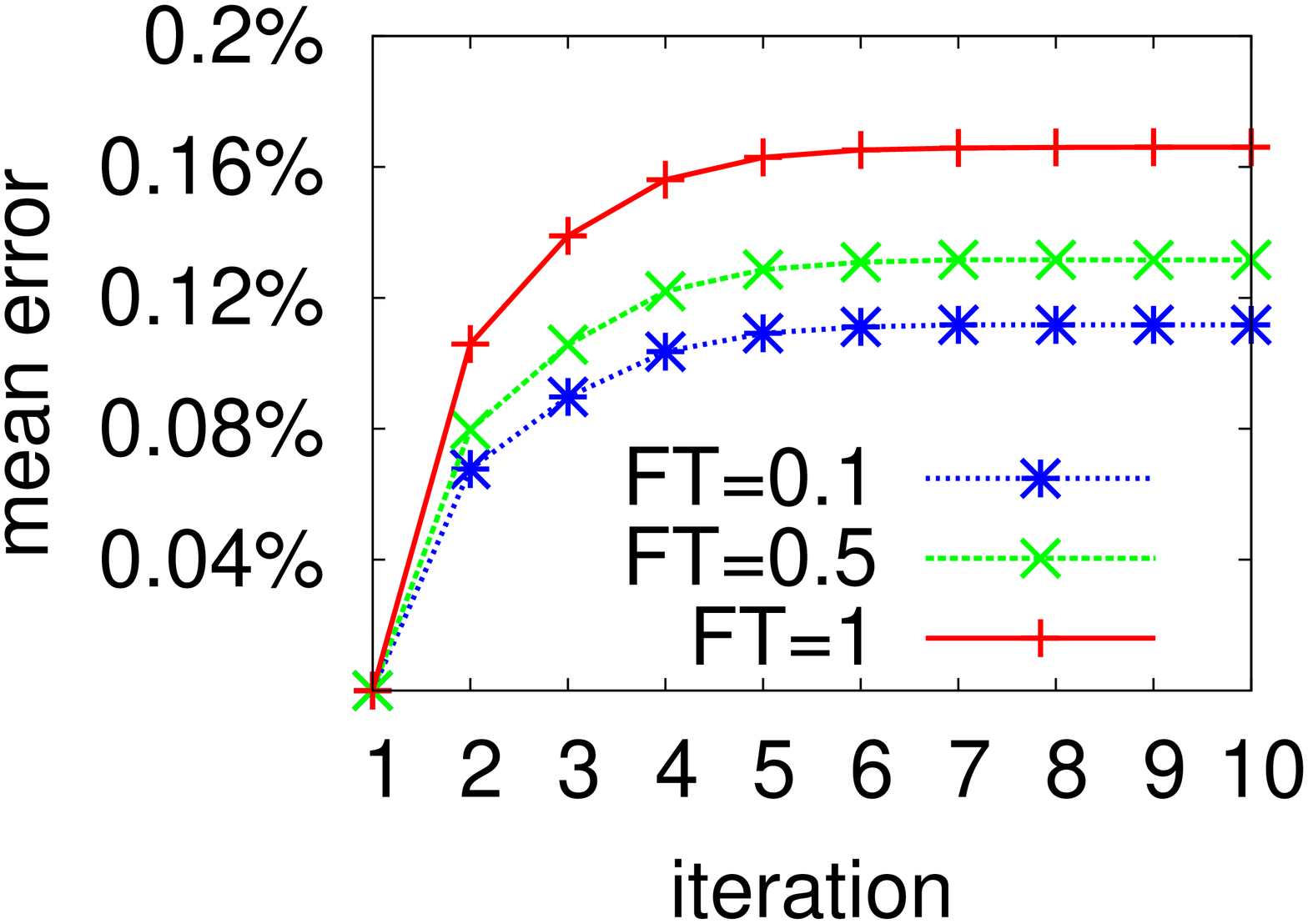}
    \label{fig:filter:error}}
    }
	\caption{Effect of change propagation control.}
	\label{fig:filter}
\end{figure}

We run PageRank on i$^2$MapReduce with 10\% changed data while varying
the change propagation filter threshold from 0.1, 0.5, to 1. (Note
that, in all previous experiments, the filter threshold is set to 1.)
Fig.~\ref{fig:filter} (a) shows the run time, while
Fig.~\ref{fig:filter} (b) shows the mean error of the kv-pairs, which
is the average relative difference from the correct value (computed
offline).

The change propagation control technique filters out the kv-pairs
whose changes are less than a given threshold.  These filtered
kv-pairs are considered very close to convergence.  As expected, the
larger the threshold, the more kv-pairs will be filtered, and the
better the run time.  On the other hand, larger threshold impacts the
result accuracy with a larger mean error.  Note that ``influential''
kv-pairs that see significant changes will hardly be filtered, and
therefore result accuracy is somewhat guaranteed.  In the experiments,
all mean errors are less than 0.2\%, which is small and acceptable.
For applications that have high accuracy requirement, users have the
option to turn off change propagation control.

In order to see the effect of change propagation control in each iteration, we show the number of propagated (non-converged) kv-pairs and the runtime per iteration with and without change propagation control. We evaluate PageRank using the ClueWeb data set. To clearly see the increasing number
of propagated kv-pairs, we randomly update only 1\% of the ClueWeb
data set, which means that there are 200,000 changed structure
kv-pairs before the incremental computation starts (iteration 0).
During the incremental processing, we record the number of propagated
kv-pairs (prop. kv-pairs) and the per-iteration-runtime after each
iteration.

\begin{figure}[htb]
	\centerline{\subfloat[Number of propagated kv-pairs]{\includegraphics[width=1.25in, angle=-90]{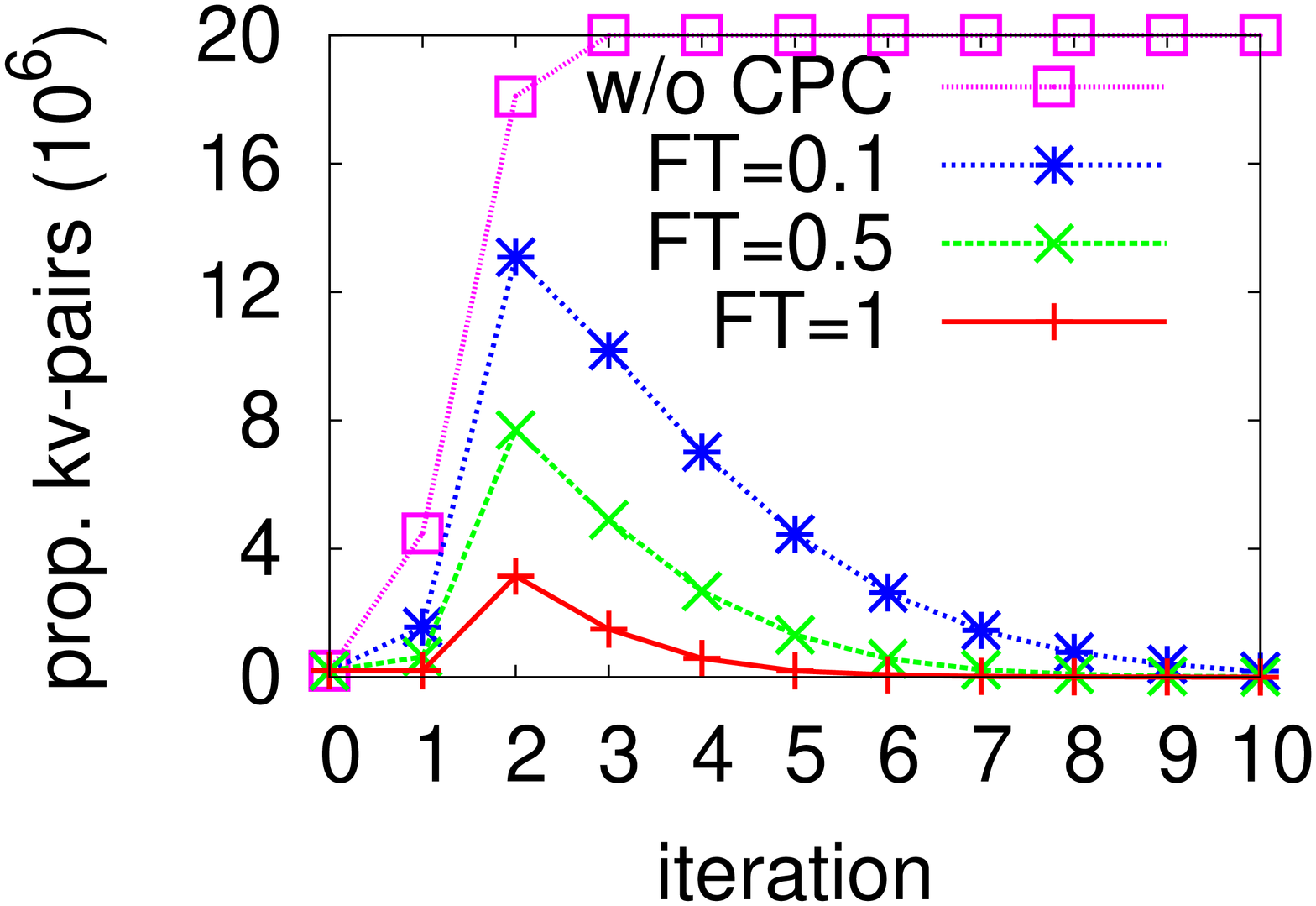}
    \label{fig:wcpc:kvs}}
	\subfloat[Runtime]{\includegraphics[width=1.25in, angle=-90]{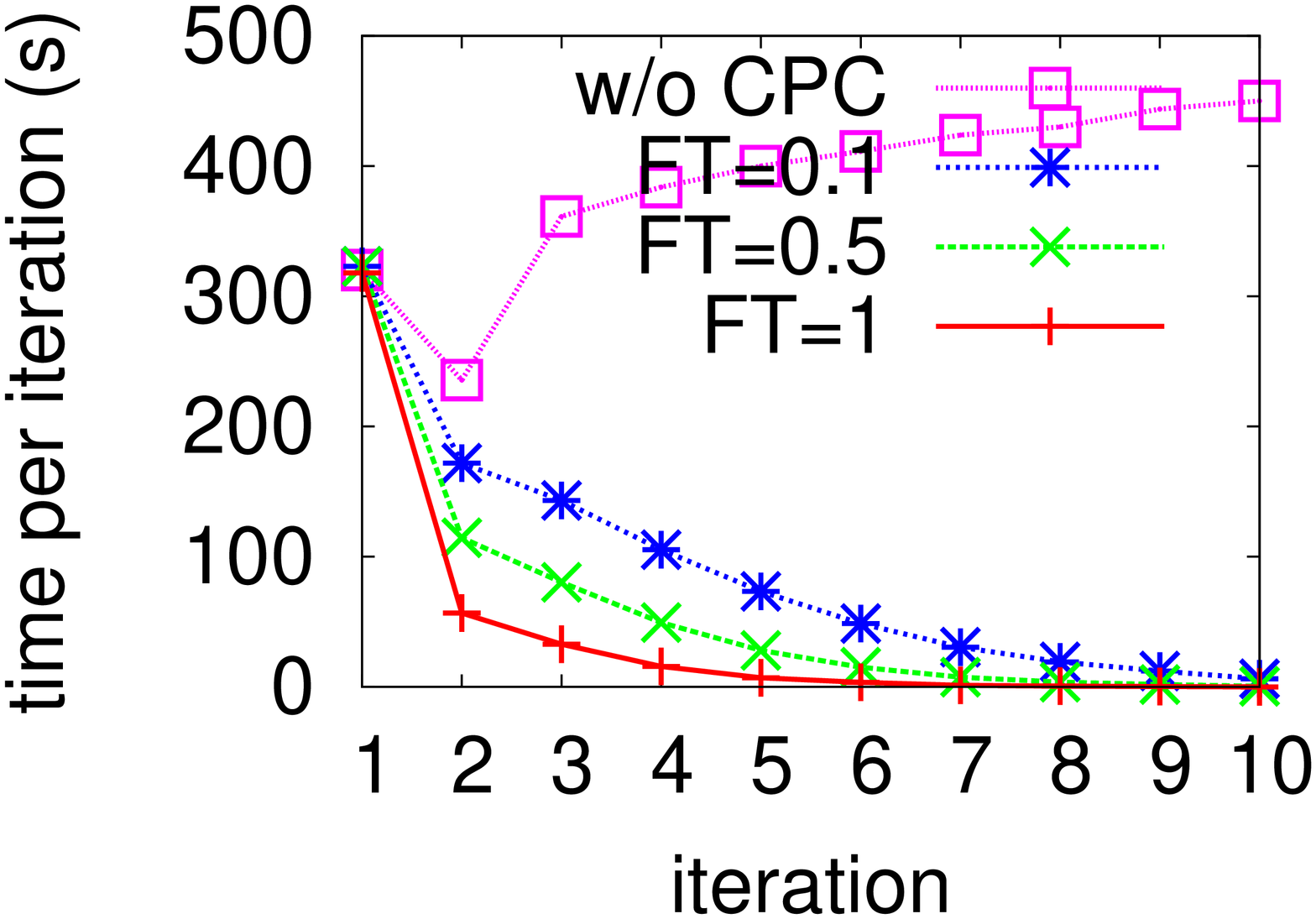}
    \label{fig:wcpc:time}}
    }
	\caption{Effect of change propagation control.}
	\label{fig:wcpc}
\end{figure}

We first run PageRank on i$^2$MapReduce without change propagation
control (i.e., w/o CPC). The number of propagated kv-pairs of each
iteration is depicted in Fig. \ref{fig:wcpc:kvs}, and the runtime of
each iteration is depicted in Fig. \ref{fig:wcpc:time}. We can see
that the changes are quickly propagated to all the kv-pairs
($20\times10^6$) after three iterations (i.e., all the Mappers and
Reducers should be re-executed). As a result, the runtime per
iteration is greatly prolonged. Further, due to the overhead of
MRBGraph maintenance, the per-iteration-runtime is steadily
increasing.  The total runtime (3859s) is just a little bit shorter
than the vanilla MapReduce (4140s). This is because that MRBGraph
maintains all the intermediate computation state, which will lead to
additional maintenance cost (accessing/update cost). If all the
Mappers/Reducers are re-executed, it is better to re-start computation
from the previously converged state with iterative processing engine
but without using MRBGraph.

We also run PageRank on i$^2$MapReduce with change propagation control varying the filter threshold (i.e., FT=1, FT=0.5, FT=0.1). Fig. \ref{fig:wcpc:kvs} depicts the number of propagated kv-pairs of each iteration, and Fig. \ref{fig:wcpc:time} depicts the runtime of each iteration. We can see that the number of non-converged kv-pairs first increases and then decreases steadily. The update of the structure data will change the previously converged result and spread the change widely in the early stage. But the incremental update will not change the converged value significantly. Consequently, the change propagation control technique will filter the kv-pairs with minor changes and reduce the per-iteration-runtime iteration by iteration. Note that, i$^2$MapReduce needs to merge the delta MRBGraph and the preserved MRBGraph in the first iteration, so the runtime of the first iteration is longer.

\subsection{HaLoop vs. iterMR}
\label{sec:vshaloop}

As mentioned in Section \ref{subsec:iterapi},
HaLoop~\cite{haloop} is one of the recent efforts that aim to improve
iterative processing on MapReduce. The other efforts include
Twister~\cite{Ekanayake:2010:TRI:1851476.1851593} and
iMapReduce~\cite{imapreduce}.  These efforts mainly focus on two
aspects: reducing job startup costs and caching structure data. Our
iterative processing engine (iterMR) also integrates these previous
optimization techniques.   Compared to HaLoop, i$^2$MapReduce can
automatically capture dependencies between structure kv-pairs and
state kv-pairs (by a user defined function Project). On the other
hand, HaLoop employs an extra MapReduce job in each iteration to join
the structure and state data.  That is, HaLoop requires two MapReduce
jobs in each iteration.

\renewcommand{\algorithmicrequire}{\textbf{Map Phase}}
\renewcommand{\algorithmicensure}{\textbf{Reduce Phase}}

\begin{algorithm}
\caption{PageRank in HaLoop}
\label{alg:prhaloop}
\begin{algorithmic}[1]
\REQUIRE1: output all inputs $<i,R_i>$ or $<i,N_i>$
\item[]
\ENSURE1: input: $<i$, $R_i|N_i>$
\FORALL{$j$ in $N_i$}
\STATE $R_{i,j}=\frac{R_i}{|N_i|}$
\STATE output $<j,R_{i,j}>$
\ENDFOR
\item[]
\REQUIRE2: output all inputs $<j,R_{i,j}>$
\item[]
\ENSURE2: input: $<j,\{R_{i,j}\}>$
\STATE $R_j=d\sum_i R_{i,j} + (1-d)$
\STATE output $<j,R_j>$
\end{algorithmic}
\end{algorithm}

We show the implementation of PageRank under HaLoop in Algorithm
\ref{alg:prhaloop}. In HaLoop, the structure and state data are
considered as two separated input data sets (i.e., map input kv-pair
is $\langle i,R_i\rangle$ or $\langle i,N_i\rangle$). We can see that
HaLoop employs an extra MapReduce job in each iteration to join the
ranking scores $<i,R_i>$ and the out-edges $<i,N_i>$ of each vertex.
HaLoop provides caching mechanism for the structure data $<i,N_i>$ in
Reduce Phase 1 to improve performance. It improves performance with
the assumption that each iteration of PageRank algorithm is
implemented in two MapReduce jobs.

In comparison, each iteration of PageRank can be implemented in a
single MapReduce job as depicted in Algorithm \ref{alg:pagerank}.  Unlike HaLoop, the structure and state data are provided
together in the input (e.g., map input kv-pair is $\langle
i,N_i|R_i\rangle$). Further, by exploring the dependencies between
structure kv-pairs and state kv-pairs, i$^2$MapReduce can
automatically join these two kinds of data, and at the same time
exploit caching optimization to further improve performance.

\subsection{Spark vs. iterMR}
\label{sec:vsspark}

In this section, we compare i$^2$MapReduce with Spark~\cite{sparknsdi}
for supporting PageRank computation.

Spark was developed to optimize large-scale interactive computation.
It uses caching techniques and operates \textbf{memory-resident}
read-only objects to improve performance.  The main abstraction in
Spark is resilient distributed dataset (RDD).  An RDD is a read-only
data set that supports only bulk processing (i.e. an operation on RDD
will be applied to each data item in the set).  Spark typically
maintains intermediate data sets across the memory of multiple
machines, and performs linkage based re-computation to recover from
failures.

Unlike disk-based systems (e.g., Hadoop, HaLoop, i$^2$MapReduce),
Spark relies on memory for fast iterative computation.  A Spark
program can separate the loop-variant state data from loop-invariant
structure data by using \texttt{partitionBy} and \texttt{join}
interfaces.  However, since RDDs are read-only, Spark will generate a
new RDD for the state data in each iteration.  Hence, Spark will

\begin{table}[!t]
    \caption{Data Sets for PageRank}
    \label{tab:dataset}
    \centering
    \footnotesize
    \begin{tabular}{c|c|c|c}
    \hline
    data set & size & \# pages & \# links \\
\hline\hline
ClueWeb-xs & 168 MB & 100,000 & 1,650,050\\
\hline
ClueWeb-s & 1.9 GB & 1,000,000 & 18,945,222\\
\hline
ClueWeb-m & 18.5 GB & 10,000,000 & 181,571,298 \\
\hline
ClueWeb-l & 36.4 GB & 20,000,000 & 365,684,186 \\
\hline
\end{tabular}
\end{table}

We use Spark 1.1.0 in our experiment.  Each Spark worker node is
configured with 2.7 GB memory (3.7 GB - 1.0 GB by default), and the
total memory capacity of the cluster is 85.2 GB.  The data sets used
in this experiment are described in Table \ref{tab:dataset}.  The web
graphs are generated based on the same approach described in Section
\ref{sec:dataset}.

\begin{figure}[t]
    \centering
        \includegraphics[width=2.9in]{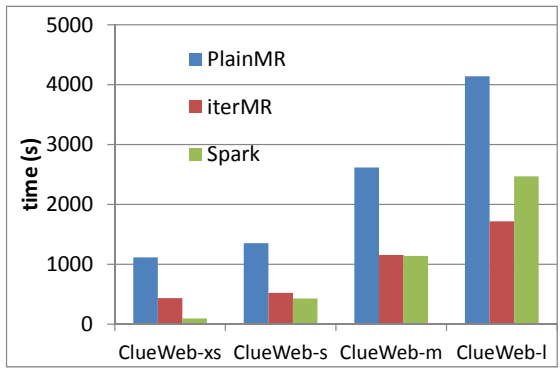}
    \caption{Run time of individual stages in PageRank.}
    \label{fig:spark}
\end{figure}

We perform PageRank on vanilla Hadoop (PlainMR), iterMR
(i$^2$MapReduce with iterative processing engine), and Spark. The
results are shown in Fig. \ref{fig:spark}. We can see that Spark is
really fast when processing small data sets (e.g., ClueWeb-xs).
However, as the input data gets larger (e.g., ClueWeb-s and
ClueWeb-m), Spark and iterMR exhibit similar performance, which is
2.5x faster than PlainMR.  However, when processing the ClueWeb-l data
set, Spark is not as good as iterMR. This is because the input data
and the intermediate data are too large, resulting degraded Spark
performance.  Therefore, the in-memory system Spark outperforms other
file-based systems if memory resource is plentiful.  However, when
processing large data set that could exhaust the memory heap space,
the performance of Spark is not satisfactory.

\subsection{Evaluation of Fault Tolerance}

\begin{figure}[t]
    \centering
        \includegraphics[width=2.4in, angle=-90]{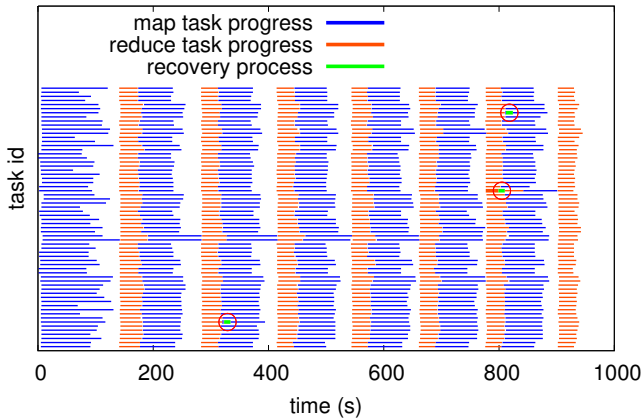}
    \caption{Fault recovery progress in i$^2$MapReduce.}
    \label{fig:fault}
\end{figure}

We test the fault recovery strategy in the context of PageRank computation. The PageRank computation is performed on the ClueWeb dataset with 32 EC2 medium instances. We configure the PageRank job with 64 prime Map tasks and 64 prime Reduce tasks. We manually and randomly inject some errors in these tasks. Figure \ref{fig:fault} depicts the execution progress of the 64 map/reduce tasks in the first 7 iterations. The length of each line indicates the runtime of each map/reduce task. We can see that there are 3 errors occurred: (1) map task 7 of iteration 3 fails at 323s; (2) reduce task 39 of iteration 6 fails at 799s; (3) map task 58 of iteration 7 fails at 812s. All the failed task can recover from failure within 12 seconds and do not impact the overall performance a lot. The failures of map task 7 and map task 58 actually do not prolong the computation process since these tasks finish earlier than the slowest tasks in a synchronization barrier.

\section{Related Work}
\label{sec:related}

\Paragraph{Iterative Processing} A number of distributed frameworks
have recently emerged for big data
processing~\cite{distgraphlab,rex,Ewen:2012:SFI:2350229.2350245,piccolo,priter,maiter}.
We discuss the frameworks that improve MapReduce.
HaLoop~\cite{haloop}, a modified version of Hadoop, improves the
efficiency of iterative computation by making the task scheduler
loop-aware and by employing caching mechanisms.
Twister~\cite{Ekanayake:2010:TRI:1851476.1851593} employs a
lightweight iterative MapReduce runtime system by logically
constructing a Reduce-to-Map loop. iMapReduce~\cite{imapreduce} supports iterative processing by
directly passing the Reduce outputs to Map and by distinguishing
variant state data from the static data. i$^2$MapReduce improves upon these previous proposals
by supporting an efficient general-purpose iterative model.

Unlike the above MapReduce-based systems, Spark \cite{sparknsdi} uses
a new programming model that is optimized for \textbf{memory-resident}
read-only objects. Spark will produce a large amount of
intermediate data in memory during iterative computation.  When input
is small, Spark exhibits much better performance than Hadoop because
of in-memory processing.  However, its performance suffers when input
and intermediate data cannot fit into memory. Our experimental results shown in Section \ref{sec:vsspark} shows that i$^2$MapReduce achieves better performance when input data is large.

Pregel \cite{1807184} follows the Bulk
Synchronous Processing (BSP) model. The computation is broken
down into a sequence of supersteps. In each superstep, a Compute
function is invoked on each vertex.  It communicates with other
vertices by sending and receiving messages and performs computation
for the current vertex.  This model can efficiently support a large
number of iterative graph algorithms. Open source implementations of Pregel include Giraph \cite{giraph}, Hama~\cite{hama}, and Pregelix \cite{pregelix}. Compared to i$^2$MapReduce, the BSP model in Pregel is quite different from the MapReduce programming paradigm. It would be interesting future work to exploit similar ideas in this paper to support incremental processing in Pregel-like systems.

\Paragraph{Incremental Processing for One-Step Application} Besides
Incoop \cite{incoop}, several recent studies aim at supporting
incremental processing for one-step applications. Stateful Bulk
Processing \cite{Logothetis:2010:SBP:1807128.1807138} addresses the
need for stateful dataflow programs. It provides a groupwise
processing operator Translate that takes state as an explicit input to
support incremental analysis. But it adopts a new programming model
that is very different from MapReduce. In addition, several research
studies \cite{incmr,Jorg:2011:IRM:2064085.2064088} support incremental
processing by task-level re-computation, but they require users to
manipulate the states on their own. In contrast, i$^2$MapReduce
exploits a fine-grain kv-pair level re-computation that are more
advantageous.

\Paragraph{Incremental Processing for Iterative Application} Naiad
\cite{Murray:2013:NTD:2517349.2522738} proposes a timely dataflow
paradigm that allows stateful computation and arbitrary nested
iterations.  To support incremental iterative computation, programmers
have to completely rewrite their MapReduce programs for Naiad.  In
comparison, we extend the widely used MapReduce model for incremental
iterative computation. Existing MapReduce programs can be slightly
changed to run on i$^2$MapReduce for incremental processing.

\section{Conclusion}
\label{sec:conclude}

We have described i$^2$MapReduce, a MapReduce-based
framework for incremental big data processing. i$^2$MapReduce combines
a fine-grain incremental engine, a general-purpose iterative model,
and a set of effective techniques for incremental iterative
computation.  Real-machine experiments show that i$^2$MapReduce can
significantly reduce the run time for refreshing big data mining
results compared to re-computation on both plain and iterative
MapReduce.

\section*{Acknowledgments}
This work was partially supported by National Natural Science Foundation of China (61300023, 61433008, 61272179), Fundamental Research Funds for the Central Universities (N141605001, N120816001), China Mobile Labs (MCM20125021), and MOE-Intel Special Fund of Information Technology (MOE-INTEL-2012-06). The second author is partially supported by the CAS Hundred Talents program and by NSFC Innovation Research Group No. 61221062.




{
\footnotesize
\bibliographystyle{abbrv}
\bibliography{references}

\begin{thebibliography}{10}

\bibitem{giraph}
Apache giraph. http://giraph.apache.org/.

\bibitem{hama}
Apache hama. https://hama.apache.org/.

\bibitem{Agrawal:1994:FAM:645920.672836}
R.~Agrawal and R.~Srikant.
\newblock Fast algorithms for mining association rules in large databases.
\newblock In {\em Proc. of VLDB '94}, pages 487--499, 1994.

\bibitem{incoop}
P.~Bhatotia, A.~Wieder, R.~Rodrigues, U.~A. Acar, and R.~Pasquin.
\newblock Incoop: Mapreduce for incremental computations.
\newblock In {\em Proc. of SOCC '11}, 2011.

\bibitem{Brin:1998:ALH:297810.297827}
S.~Brin and L.~Page.
\newblock The anatomy of a large-scale hypertextual web search engine.
\newblock {\em Comput. Netw. ISDN Syst.}, 30(1-7):107--117, Apr. 1998.

\bibitem{pregelix}
Y.~Bu, V.~Borkar, J.~Jia, M.~J. Carey, and T.~Condie.
\newblock Pregelix: Big(ger) graph analytics on a dataflow engine.
\newblock {\em PVLDB}, 8(2):161--172, 2015.

\bibitem{haloop}
Y.~Bu, B.~Howe, M.~Balazinska, and M.~D. Ernst.
\newblock Haloop: efficient iterative data processing on large clusters.
\newblock {\em PVLDB}, 3(1-2):285--296, 2010.

\bibitem{ilprints376}
J.~Cho and H.~Garcia-Molina.
\newblock The evolution of the web and implications for an incremental crawler.
\newblock Technical Report 1999-22, Stanford InfoLab, 1999.

\bibitem{1251264}
J.~Dean and S.~Ghemawat.
\newblock Mapreduce: simplified data processing on large clusters.
\newblock In {\em Proc. of OSDI '04}, 2004.

\bibitem{Ekanayake:2010:TRI:1851476.1851593}
J.~Ekanayake, H.~Li, B.~Zhang, T.~Gunarathne, S.-H. Bae, J.~Qiu, and G.~Fox.
\newblock Twister: a runtime for iterative mapreduce.
\newblock In {\em Proc. of MAPREDUCE '10}, 2010.

\bibitem{Ewen:2012:SFI:2350229.2350245}
S.~Ewen, K.~Tzoumas, M.~Kaufmann, and V.~Markl.
\newblock Spinning fast iterative data flows.
\newblock {\em PVLDB}, 5(11):1268--1279, 2012.

\bibitem{Jorg:2011:IRM:2064085.2064088}
T.~J\"{o}rg, R.~Parvizi, H.~Yong, and S.~Dessloch.
\newblock Incremental recomputations in mapreduce.
\newblock In {\em Proc. of CloudDB '11}, 2011.

\bibitem{5360248}
U.~Kang, C.~Tsourakakis, and C.~Faloutsos.
\newblock Pegasus: A peta-scale graph mining system implementation and
  observations.
\newblock In {\em Proc. of ICDM '09}, pages 229--238, 2009.

\bibitem{Kwon:2012:SMS:2213836.2213840}
Y.~Kwon, M.~Balazinska, B.~Howe, and J.~Rolia.
\newblock Skewtune: Mitigating skew in mapreduce applications.
\newblock In {\em Proc. of SIGMOD '12}, pages 25--36, 2012.

\bibitem{lloyd1982least}
S.~Lloyd.
\newblock Least squares quantization in pcm.
\newblock {\em Information Theory, IEEE Transactions on}, 28(2):129--137, 1982.

\bibitem{Logothetis:2010:SBP:1807128.1807138}
D.~Logothetis, C.~Olston, B.~Reed, K.~C. Webb, and K.~Yocum.
\newblock Stateful bulk processing for incremental analytics.
\newblock In {\em Proc. of SOCC '10}, 2010.

\bibitem{distgraphlab}
Y.~Low, D.~Bickson, J.~Gonzalez, C.~Guestrin, A.~Kyrola, and J.~M. Hellerstein.
\newblock Distributed graphlab: a framework for machine learning and data
  mining in the cloud.
\newblock {\em PVLDB}, 5(8):716--727, 2012.

\bibitem{1807184}
G.~Malewicz, M.~H. Austern, A.~J. Bik, J.~C. Dehnert, I.~Horn, N.~Leiser, and
  G.~Czajkowski.
\newblock Pregel: a system for large-scale graph processing.
\newblock In {\em Proc. of SIGMOD '10}, 2010.

\bibitem{rex}
S.~R. Mihaylov, Z.~G. Ives, and S.~Guha.
\newblock Rex: recursive, delta-based data-centric computation.
\newblock {\em PVLDB}, 5(11):1280--1291, 2012.

\bibitem{Murray:2013:NTD:2517349.2522738}
D.~G. Murray, F.~McSherry, R.~Isaacs, M.~Isard, P.~Barham, and M.~Abadi.
\newblock Naiad: A timely dataflow system.
\newblock In {\em Proc. of SOSP '13}, pages 439--455, 2013.

\bibitem{olston2010web}
C.~Olston and M.~Najork.
\newblock Web crawling.
\newblock {\em Foundations and Trends in Information Retrieval}, 4(3):175--246,
  2010.

\bibitem{Peng:2010:LIP:1924943.1924961}
D.~Peng and F.~Dabek.
\newblock Large-scale incremental processing using distributed transactions and
  notifications.
\newblock In {\em Proc. of OSDI '10}, pages 1--15, 2010.

\bibitem{piccolo}
R.~Power and J.~Li.
\newblock Piccolo: Building fast, distributed programs with partitioned tables.
\newblock In {\em Proc. of OSDI'10}, pages 1--14, 2010.

\bibitem{incmr}
C.~Yan, X.~Yang, Z.~Yu, M.~Li, and X.~Li.
\newblock Incmr: Incremental data processing based on mapreduce.
\newblock In {\em Proc. of CLOUD '12}, 2012.

\bibitem{sparknsdi}
M.~Zaharia, M.~Chowdhury, T.~Das, A.~Dave, J.~Ma, M.~McCauley, M.~J. Franklin,
  S.~Shenker, and I.~Stoica.
\newblock Resilient distributed datasets: A fault-tolerant abstraction for.
  in-memory cluster computing.
\newblock In {\em Proc. of NSDI '12}, 2012.

\bibitem{priter}
Y.~Zhang, Q.~Gao, L.~Gao, and C.~Wang.
\newblock Priter: A distributed framework for prioritized iterative
  computations.
\newblock In {\em Proc. of SOCC '11}, 2011.

\bibitem{maiter}
Y.~Zhang, Q.~Gao, L.~Gao, and C.~Wang.
\newblock Accelerate large-scale iterative computation through asynchronous
  accumulative updates.
\newblock In {\em Proc. of ScienceCloud '12}, 2012.

\bibitem{imapreduce}
Y.~Zhang, Q.~Gao, L.~Gao, and C.~Wang.
\newblock imapreduce: A distributed computing framework for iterative
  computation.
\newblock {\em J. Grid Comput.}, 10(1), 2012.

\end{thebibliography}
}

\end{document}